\documentclass[11pt,a4paper]{article}

\usepackage{amsmath,amssymb}
\usepackage{graphicx}
\usepackage{booktabs}
\usepackage[margin=2.5cm]{geometry}
\usepackage{subcaption}
\usepackage{placeins}
\usepackage{algorithm}
\usepackage{algpseudocode}
\usepackage{float}
\usepackage{natbib}
\usepackage[hidelinks]{hyperref}
\usepackage[normalem]{ulem}

\algrenewcommand\algorithmicrequire{\textbf{Input:}}
\algrenewcommand\algorithmicensure{\textbf{Output:}}

\begin{document}

\begin{titlepage}
\setlength{\parindent}{0pt}

\vspace*{2cm}

\begin{center}
\LARGE\bfseries Randomized Neural Networks for estimation of exposure profiles and Credit Valuation Adjustment (CVA) for American Equity Options
\end{center}

\vspace{1.5cm}

\begin{center}
\large
Isidro Moroso Varona$^{1}$, Jakub Michańków$^{2}$, Paweł Sakowski$^{2}$\\[0.5cm]
\vspace{1.5cm}
\small
$^{1}$Faculty of Economic Sciences, University of Warsaw, \texttt{i.morosovaro@student.uw.edu.pl }\\
$^{2}$Department of Quantitative Finance and Machine Learning,\\Faculty of Economic Sciences, University of Warsaw, \texttt{j.michankow@uw.edu.pl}, \texttt{p.sakowski@uw.edu.pl}\\
[0.5cm]

\end{center}

\vfill

\begin{center}
October 2025
\end{center}

\vspace{1cm}


\end{titlepage}

\newpage
\section*{\centering\normalsize\textbf{Abstract} }

\begin{center}
\parbox{1\textwidth}{
This paper studies the use of randomized neural networks for the estimation of exposure profiles and unilateral CVA of American options within a Monte Carlo framework. The analysis is carried out separately under both Black-Scholes and Heston dynamics, combining American option valuation, expected exposure and potential future exposure estimation, and unilateral CVA calculation with portfolio netting effects. 

The numerical experiment compares this approach with the classical Least-Squares Monte Carlo (LSM) used as a benchmark in both low-dimensional single-asset and high-dimensional multi-asset scenarios, and also includes a path convergence test and a sensitivity analysis. The results show that the randomized feedforward neural network approach preserves convergence to the LSM benchmark when it is extended from pricing to exposure and CVA estimation, while its main advantage appears in high-dimensional problems, where it scales more efficiently and leads to lower computational cost. 

These results support the use of randomized neural networks as a useful alternative for exposure and CVA estimation in high-dimensional American-style options.
}
\end{center}

\bigskip
\section*{\centering\normalsize\textbf Keywords:}

\begin{center}
Credit Valuation Adjustment (CVA), Counterparty Credit Risk (CCR), exposure profiles, Expected Exposure (EE), Potential Future Exposure (PFE), pricing, American options, Least-Squares Monte Carlo (LSM), randomized neural networks, Randomized Least-Squares Monte Carlo (RLSM)

\vspace{2cm}
Thematic classification\\
C4, C14, C45, C53, C58, G13

\end{center}

\newpage

\tableofcontents
\newpage

\setlength{\parindent}{17pt}
\setlength{\parskip}{6pt}
\clearpage
\section*{Introduction} \label{introduction}
\addcontentsline{toc}{section}{Introduction}

The exposure profiles of derivatives are used for the calculation of Credit Valuation Adjustment (CVA) and other Counterparty Credit Risk (CCR) measures. However, for the estimation of exposure profiles in early exercise products it is required to account for the optimal stopping policy.
The motivation of this paper is to extend the randomized neural network model proposed by \citep{HerreraKrachRuyssenTeichmann2023} from pricing to the estimation of the exposure profiles and Credit Valuation Adjustment (CVA) calculations for American options. This extension means to apply the model to a more demanding problem since the  exposure curves depend on both the option future prices and exercise dates.

The ideal scope of application of this model is in high-dimensional products, where the dimension refers to the number of underlying assets or risk factors driving the contract, since in that setting the classical models become harder to handle and the randomized neural network approach of \citep{HerreraKrachRuyssenTeichmann2023} may offer a more efficient and reliable alternative. The question studied here is therefore whether that framework still converges to the same exposure, price, and CVA values as the benchmark once it is used for exposure and CVA estimation.

The main objective is to study whether the randomized neural network framework proposed by \citep{HerreraKrachRuyssenTeichmann2023} can be extended from pricing to the estimation of exposure profiles and unilateral CVA for American options, while still converging to the same benchmark values.

To achieve this main objective is to include to develop a Monte Carlo framework for valuing American options under Black-Scholes and Heston dynamics, implement both the classical Least-Squares Monte Carlo method as a benchmark and the randomized neural network approach, generate exposure profiles from the simulated option values, and estimate unilateral CVA with netting effects for an options portfolio from those profiles. The comparison between both methods then will focus on the convergence of the results and analysis of the computational costs.

To check whether these objectives are achieved, the research questions will be answered at the end of the article once the numerical results have been analyzed. These research questions are the following:
\begin{enumerate}
\item Does the randomized neural network approach converge to the same exposure profiles as the LSM benchmark?
\item Does the randomized neural network approach scale more efficiently than LSM as the dimension of the problem increases?
\item Does this convergence remain once the exposure profiles are used to compute unilateral CVA, including portfolio aggregation with netting effects?
\end{enumerate}

The article is structured as follows. Chapter \ref{literature review} presents the literature review and introduces the main references and concepts that motivate the work. Chapter \ref{theoretical background} develops the theoretical background, covering counterparty credit risk, exposure profiles, American-style options, Monte Carlo valuation, Least-Squares Monte Carlo and randomized neural networks for optimal stopping. Chapter \ref{methodology} describes the methodology used in the article, including the overall simulation framework, the path generation models, the LSM and Randomized Least-Squares Monte Carlo (RLSM) implementations, and the CVA framework. Chapter \ref{experiment design} sets out the experimental design, specifying the products, model setup, Monte Carlo configuration, reference values and evaluation metrics used in the numerical analysis. Chapter \ref{experiment results} presents and discusses the numerical results for vanilla American options and high-dimensional max-call options, together with the path convergence and sensitivity analyses. Chapter \ref{cva case study} contains the CVA case study, including both trade-level and portfolio-level results. Finally, the Chapter of \hyperref[conclusions]{conclusions} summarizes the main findings and discusses the main limitations of the work and possible further research.

\section{Literature Review} \label{literature review}

Future exposure estimation became a major issue in derivatives valuation due to 
the quick development of counterparty credit risk modeling after the global 
financial crisis of 2008. From then on, the CVA literature expanded alongside the larger 
xVA literature, with a focus on the precise calculation of expected exposure profiles, 
portfolio netting and collateral effects, and credit-adjusted values \citep{Gregory2015,CesariAquilinaCharpillon2009}. 
However, the issue becomes more complex when the contracts under consideration have 
early-exercise features because exposure and the optimal stopping rule became inherent. 
This is the exact point at which the literature on American and Bermudan option valuation 
becomes directly applicable to counterparty risk applications \citep{CesariAquilinaCharpillon2009,BrigoMoriniPallavicini2013}.

Traditional Monte Carlo methods were highly inefficient for valuing these types of products, since at each exercise date the option value depends on the comparison between the immediate payoff and the continuation value, which represents the expected value of keeping the option alive rather than exercising it immediately, and a direct Monte Carlo implementation would require nested simulations to estimate that continuation value, making the computational cost grow exponentially with the number of dates. As an alternative, regression-based and dynamic programming
aproximations were proposed by \citep{Carriere1996}, later by \citep{TsitsiklisVanRoy1997,TsitsiklisVanRoy2001}, and finally 
\citep{LongstaffSchwartz2001} introduced the least-squares Monte Carlo method (LSM), that quickly became
an industry standard.

Later on, the same regression-based logic expanded to exposure and counterparty credit
risk calculations, where Longstaff-Schwartz simulation, is also known as American Monte Carlo (AMC), 
became frequently used. In that context, the method is not only used to estimate one
price value but also for estimating exposure profiles based on future values and optimal
stopping rule. This use of LSM/AMC for risk estimation purposes and credit valuation adjustment, is well
documented within xVA literature and in later comparisons, focus more on estimating exposures
rather than pricing \citep{Gregory2015,CesariAquilinaCharpillon2009,Cortis2019}.

As a continuation of \citep{LongstaffSchwartz2001} method, a broad literature on alternative optimal stopping models has
emerged, most of them motivated by pricing rather than by exposure
applications. Among these models, the most influential lines are stochastic mesh methods \citep{BroadieGlasserman2004},
dual and primal-dual formulations \citep{Rogers2002,AndersenBroadie2004,BroadieCao2008}, and bundling-based approaches\citep{JainOosterlee2015}.

Other more recent contributions in the literature focus on the limitations of LSM for high dimensional
products that affects the model stability and computational costs. Among these alternatives to traditional
regression-based models, different Machine Learning approaches that uses the same risk factors as features are proposed, some of them include deep-learning-based 
stopping policies, neural networks continuation-value estimation, neural approaches designed directly for exposure profiles, 
and reinforcement-learning alternatives
\citep{BeckerCheriditoJentzen2019,LapeyreLelong2021,AnderssonOosterlee2021,HerreraKrachRuyssenTeichmann2023}. 
In other words, the literature gradually moves from classical regression toward
richer approximation architectures once the dimensionality or the complexity of
the product makes standard polynomial bases too restrictive or too costly.

Among these new alternatives, randomized neural network methods occupy an
interesting middle ground, since they retain a relatively simple structure,
avoid the cost of fully training a deep network using back-propagation, 
and still provide a nonlinear approximation of the continuation value. 
This is the idea proposed by \citep{HerreraKrachRuyssenTeichmann2023}, who study optimal stopping through randomized
neural networks and compare several possible approaches within that framework. This article is presented
as a continuation of that literature, but with a different focus, since 
our interest is not only to examine whether randomized neural networks
can price American options, but also whether they can produce accurate
exposure profiles and CVA estimates in the high-dimensional
setting where the LSM benchmark becomes exponentially more expensive. A related but distinct approach modifies the regression step itself rather than the network architecture: \citep{huo2025finite} introduces a finite-difference solution ansatz within least-squares Monte Carlo, improving the accuracy of the continuation-value estimate without departing from the LSM framework.

\clearpage
\section{Theoretical Background} \label{theoretical background}

\subsection{Counterparty Credit Risk and Credit Valuation Adjustment}

\subsubsection{Counterparty Credit Risk}

Counterparty credit risk appears when one party to a derivatives contract may suffer a loss because the other party fails to perform its obligations before the transaction has been fully settled. In derivatives markets this risk has a particular shape, since the exposure is not fixed from the start in the same way as for a conventional loan or bond position. The future value of a derivative depends on market movements, on the contractual structure of the trade and, in some cases, on optional features embedded in the payoff, so the amount that is effectively at risk changes through time rather than remaining known in advance \citep{Gregory2010CCR}.

In the over-the-counter market, derivatives contracts are typically negotiated bilaterally and exposures are managed at the level of a counterparty relationship rather than as a list of isolated trades. The loss depends on the value of the trade or portfolio at the time of default, on whether contractual netting can be enforced, and on whether collateral has already reduced part of the exposure. In that sense, counterparty credit risk in derivatives is usually understood as a combination of credit risk and market risk, because the final exposure depends both on the credit quality of the counterparty and on the future market value of the position \citep{Gregory2010CCR,BISCCRBasel2018}.

Seen from that perspective, counterparty risk is broader than a simple question of solvency, as it is connected to valuation, portfolio aggregation and risk management at the same time, which is why it became such a central issue in modern derivatives practice. Even before the financial crisis, banks were already aware of this risk, but its scale and speed were often underestimated, but once the major financial institutions began to fail or to experience stress problems, it became clear that counterparty risk was not a small correction around an otherwise stable pricing framework, but something capable of changing the economic value of a derivatives book in a material way \citep{Gregory2010CCR,CesariAquilinaCharpillon2009}.

During the global financial crisis of 2007-2009, events like the default of Lehman Brothers, the stress around AIG and the broader disruption of OTC derivatives markets showed that losses linked to derivatives portfolios could arise not only because a counterparty actually defaulted, but also because the market reassessed the counterparty's credit quality and repriced that risk well before default. From that point on, counterparty credit risk was no longer treated as a specialized issue at the edge of derivatives valuation, but as one of its central components \citep{Gregory2010CCR,Gregory2015}.

\subsubsection{Basel III}

The Basel framework provides one of the main international references for bank regulation and supervision, and it did not appear in a single step. The Basel Committee on Banking Supervision was created in 1974, after a period of serious disruption in international banking markets, and over time its standards were revised and expanded as the structure of financial risk became more complex. Basel I, introduced in 1988, established the first internationally agreed capital framework, while Basel II, published in 2004, developed a broader and more risk-sensitive architecture, and Basel III emerged later as the post-crisis response to weaknesses that had become evident during the global financial crisis, especially in the treatment of capital, liquidity and counterparty-related risks \citep{BISBaselHistory}.

Under Basel III, counterparty credit risk received a much more explicit treatment, and the regulatory framework distinguished more clearly between the risk that a counterparty defaults and the losses generated by changes in credit valuation adjustment. The post-crisis reforms introduced a specific capital charge for CVA risk, largely because a substantial share of the counterparty-related losses observed during the crisis came from CVA mark-to-market losses rather than from outright defaults alone \citep{BCBSBilateralCCR2011,BISCCRBasel2018}. For the purposes of this article, this regulatory background matters mainly as context, since it helps explain why exposure modelling and CVA estimation became much more relevant from both a pricing and a risk-management perspective.

At the same time, the Basel treatment also showed that counterparty risk in derivatives cannot be treated as a static credit problem, since it depends on the future evolution of the portfolio value, on the way exposures are aggregated, and on whether those exposures can be estimated with enough accuracy for valuation and risk management, which is why simulation-based methods, and later machine-learning approximations, became more relevant after the crisis.

\subsubsection{The xVA Framework}

The broader response of practice and regulation to these issues is usually described under the term xVA. In general terms, xVA refers to the family of Valuation Adjustments applied to the clean or risk-free value of a derivatives portfolio in order to reflect credit, funding, collateral, capital and margin effects that are not captured by a purely risk-free pricing framework \citep{Gregory2015}. The notation is intentionally broad, since the letter x stands for several possible adjustments that are part of the same family.

Within that family, the most common components include credit valuation adjustment (CVA), debt valuation adjustment (DVA), funding valuation adjustment (FVA), collateral valuation adjustment (ColVA), margin valuation adjustment (MVA) and capital valuation adjustment (KVA). Their exact implementation may vary across institutions, and in practice some of them interact quite closely, but the general picture is that modern derivatives valuation no longer rests only on a clean price under idealised assumptions, but must also incorporate the economic effects of credit risk, funding conditions, collateral agreements, margin requirements and regulatory capital \citep{Gregory2015}.

During this article, our work will focus only in CVA as it's the most popular one and sufficient for analysing the effects of exposure estimations in early-exercise derivatives. However, the other valuation adjustments remain conceptually relevant, since they explain why CVA belongs to a wider post-crisis framework.

\subsubsection{Credit Valuation Adjustment (CVA)}

At an intuitive level, CVA may be understood as the reduction that must be applied to the clean value of a derivatives position in order to reflect the possibility that the counterparty defaults before all contractual cash flows have been exchanged. In other words, CVA represents the expected loss generated by future positive exposure to the counterparty \citep{Gregory2010CCR,Gregory2015}. This interpretation is standard in the counterparty risk literature and provides the main link between credit modelling and future exposure modelling.

In practice, the same idea is often expressed through the relationship
\begin{equation}
\mathrm{Risky\ value} = \mathrm{Risk\mbox{-}free\ value} - \mathrm{CVA},
\end{equation}
which makes clear that the adjustment is a deduction from the clean valuation \citep{Gregory2015}. A general representation of CVA, written in the form used for computation, is
\begin{equation}
\mathrm{CVA} = \sum_{i=1}^{m} \mathrm{LGD} \times \mathrm{EE}(t_i) \times \mathrm{PD}(t_{i-1}, t_i),
\end{equation}
where $\mathrm{LGD}$ denotes loss given default, $\mathrm{EE}(t_i)$ the discounted expected exposure at future date $t_i$, and $\mathrm{PD}(t_{i-1}, t_i)$ the marginal default probability over the interval $(t_{i-1}, t_i)$ \citep{Gregory2015}. Written in this form, it becomes very clear why the rest of the chapter will move toward exposure profiles, since exposure is not an accessory quantity here but one of the two main ingredients of CVA itself.

CVA brings together two parts of the same problem, on one side there is the future exposure generated by the trade or portfolio, which changes through time and with market conditions, and on the other side there is the credit component, meaning the possibility of default and the fact that even after default part of the claim may still be recovered. This is why survival probabilities, default probabilities, hazard rates and recovery rates appear so naturally in the CVA framework, even if they are not always modelled with the same level of detail in every application \citep{Gregory2010CCR}, and in the end the expected loss depends both on the size of the positive exposure and on the probability that default takes place before the position is settled.

At this stage it is also worth distinguishing between unilateral and bilateral CVA, in the unilateral case only the default of the counterparty is considered, while in the bilateral case the institution's own default also enters the valuation through the corresponding debt valuation adjustment \citep{Gregory2015,CesariAquilinaCharpillon2009}. Since this article is concerned with the counterparty side of the problem, the empirical chapters focus on unilateral CVA.

Once CVA has been introduced at this general level, the natural next step is to look more carefully at the exposure side of the problem. The following subsection therefore turns to the definition of exposure profiles, their evolution through time, and the role of netting when several trades are considered together.

\subsection{Exposure Profiles, Netting and Default}

Once CVA has been introduced as an expected loss, the next step is to describe the components that are used into its calculation. From a theoretical point of view, this means looking at the exposure profile of a trade or portfolio, the way this profile is modified by netting and collateral agreements, and the credit-side quantities that later interact with exposure inside the CVA formula. The purpose of this subsection is therefore to describe the main concepts that will be used later in the methodology and in the case study.

\subsubsection{Exposure Profiles}

In counterparty risk, exposure refers to the positive value of a transaction or portfolio from the point of view of the institution. If the mark-to-market value, denoted by $\mathrm{MtM}$, is negative, then the institution owes value to the counterparty and there is no credit exposure on that side. This asymmetry can be written as
\begin{equation}
\mathrm{Exposure} = \max(\mathrm{MtM},0) = \mathrm{MtM}^{+}.
\end{equation}
\citep{Gregory2010CCR}. This simple definition already shows that exposure is not the same as the raw value of the contract, because for counterparty loss, only the positive part is considered.

Since the mark-to-market evolves through time and differs across future scenarios, exposure is described as a profile rather than as a single number. The most natural quantity to introduce first is expected exposure ($EE$), since this gives the average positive exposure ($E$) at a given future date. Using the previous definition of exposure, it may be written in the natural form
\begin{equation}
EE(t)=\mathbb{E}[E(t)].
\end{equation}
This makes expected exposure the main metric of the exposure profile used later in CVA calculations. Expected positive exposure (EPE) is closely related, and is usually understood as the average of the expected exposure profile through time, while expected negative exposure (ENE) gives the corresponding profile on the negative side and is mainly relevant in bilateral settings \citep{Gregory2010CCR}.

A second measure that is widely used in practice is Potential Future Exposure (PFE) at time $t$, which focuses on tail risk rather than on the average across Monte Carlo paths. \citep{CesariAquilinaCharpillon2009} define it as
\begin{equation}
PFE_{\alpha,t}=q_{\alpha,t}=\inf\{x:\mathbb{P}(V_t\leq x)\geq \alpha\},
\end{equation}

where $\alpha$ is the chosen confidence level and $V_t$ denotes the future value distribution at time $t$. PFE answers the question of how large the exposure could become at a given future date under a high-confidence scenario. This is why EE and PFE are complementary as one captures the average profile, while the other gives information about the upper (or lower) tail of the exposure distribution.

Other exposure measures that appear in the literature, although they are not used during this article, are expected mark-to-market, which is useful as a reference quantity but does not isolate counterparty loss, expected shortfall, which may be used when more information is needed about severe tail events, and effective EPE, which is especially relevant in regulatory applications \citep{Gregory2010CCR}.

\subsubsection{Netting and Collateral}

Exposure is rarely assessed trade by trade in isolation, because in practice counterparties usually face each other through portfolios of transactions, and the relevant object is therefore the exposure of the netting set rather than that of each individual trade. This is where netting becomes crucial, since if several transactions with the same counterparty are covered by a legally enforceable netting agreement, positive and negative values may offset each other in the event of default, so the relevant loss is based on the net value of the portfolio rather than on the simple sum of standalone positive exposures \citep{Gregory2010CCR,CesariAquilinaCharpillon2009}.

From a quantitative point of view, this means that portfolio aggregation cannot in general be done by adding trade-level exposure summaries one by one. In a simulation framework, aggregation has to be carried out scenario by scenario and time step by time step, preserving the joint evolution of portfolio values \citep{CesariAquilinaCharpillon2009}.

Collateral plays a related but different role, while netting reduces exposure by offsetting positive and negative values, collateral reduces the unsecured amount that remains after that offsetting has taken place. In practice, collateral agreements specify how collateral is posted, how often margin calls are made, which thresholds apply, and what types of assets can be used \citep{Gregory2010CCR,CesariAquilinaCharpillon2009}. In an ideal situation collateral may reduce counterparty exposure substantially, but in practice it does not eliminate the risk completely, since thresholds, timing effects, disputes, rehypothecation and the margin period of risk may still leave residual exposure.

\subsubsection{Default Probability, Hazard Rate and Recovery}

As we saw in the previous section CVA also depends on the credit quality of the counterparty. The default probability measures the likelihood that a default occurs over a given time interval, while survival probability measures the probability that no default occurs up to a given horizon \citep{Gregory2010CCR}. These two notions describe the same process from complementary points of view and are part of the standard representation of credit risk over time.

In reduced-form models, this term structure of default risk is often represented through a hazard rate, or default intensity, understood as an instantaneous default probability \citep{Gregory2010CCR}. This representation is useful because it describes default risk locally through time and links survival probabilities with marginal default probabilities.

Finally, recovery rate refers to the fraction of the claim that is expected to be received after default, while loss given default represents the non-recovered part. In the notation used in the literature, this relation is written as $LGD = 100\% - Rec$ \citep{Gregory2015}. Together with the exposure profile described above, default probability, hazard rate and recovery complete the theoretical structure required to estimate the CVA.

\subsection{American-Style Options}

\subsubsection{American and Bermudan Options}

An option gives its holder the right, but not the obligation, to buy or sell an underlying asset under pre-specified conditions. In the standard European case, that right can be exercised only at maturity. However, American and Bermudan options differ from this benchmark because they allow exercise before the final date \citep{Hull2003}.

An American option can be exercised at any time up to maturity, while a Bermudan option can be exercised only on a finite set of predetermined dates. In that sense, the Bermudan contract lies between the European and the American one, since it includes early exercise but only at specific times rather than continuously \citep{Hull2003}, and this distinction is common in practice, especially in contracts whose structure naturally creates a discrete exercise schedule.

In both cases there is an exercise feature that is absent from the European contract, since the holder does not have to wait until the final maturity date, and this changes both the nature of the contract and the way it is valued.

\subsubsection{Max-Call Options}

Options may depend on several underlying assets rather than on a single one. In that setting, one common class of contracts is given by options whose payoff depends on the maximum of the underlying asset values. In the notation used by \citep{Glasserman2004} for options on the maximum of several assets, the payoff may be written as
\begin{equation}
\left(\max\{c_1S_1(T),c_2S_2(T),\ldots,c_dS_d(T)\}-K\right)^+
\end{equation}

where $S_i(T)$ denotes the price of the $i$-th underlying asset at maturity $T$, $c_i$ denotes the scaling weight of the asset $i$, and $K$ is the strike price. This type of payoff is frequently used in literature to represent high-dimensional scenarios for American derivatives \citep{LongstaffSchwartz2001,Glasserman2004,LapeyreLelong2021,BeckerCheriditoJentzen2019}.  \citep{LongstaffSchwartz2001}, for example, include an American option on the maximum of five risky assets as one of their benchmark examples. For that reason, the max-call provides a natural product for the present work as it is simple at the level of payoff definition, but it becomes increasingly computationally expensive as the number of underlying assets grows.

\subsection{Monte Carlo Valuation and the Early-Exercise Problem}

\subsubsection{Monte Carlo Valuation of Derivatives}

Monte Carlo methods became one of the main numerical tools in derivatives pricing because they make it possible to value contracts through simulated paths of the underlying risk factors rather than through a full discretization of the state space. In \citep{Boyle1977}, option valuation is already framed in terms of simulating the process followed by the underlying asset and using risk-neutral valuation to obtain the option price. Since then, Monte Carlo methods have become standard in settings where the payoff depends on several sources of uncertainty or on the evolution of the path rather than on a single terminal value \citep{Glasserman2004}.

Under the risk-neutral approach, the valuation is written as an expected discounted payoff. In the simple European call example used by \citep{Glasserman2004}, this takes the form
\begin{equation}
E[e^{-rT}(S(T)-K)^+].
\end{equation} 
The same principle extends beyond this basic case. Once the relevant stochastic model has been specified, Monte Carlo valuation proceeds by simulating many paths, computing the discounted payoff on each path, and averaging across simulations.

This framework is especially useful when the derivative is path dependent or when its value depends on many factors at the same time as the simulation often remains feasible even when tree-based or finite-difference methods become difficult to use because the dimension of the problem is too large or the payoff structure is too irregular \citep{Glasserman2004,LongstaffSchwartz2001}. For this reason, Monte Carlo provides a natural starting point for the products considered in this article.

\subsubsection{Early Exercise in Monte Carlo Pricing}

The situation changes once the early exercise optionality is introduced. For a European options, simulating paths up to maturity and averaging the discounted terminal payoff is enough. However, for American or Bermudan option, the valuation must account for the fact that the holder may exercise before the final date, so at each exercise time, the contract value is determined by the comparison between immediate exercise and continuation. The discounted value is written through the recursion
\begin{equation}
V_m(x)=h_m(x), \qquad
V_{i-1}(x)=\max\{h_{i-1}(x),E[V_i(X_i)\mid X_{i-1}=x]\}.
\end{equation}
\citep{Glasserman2004}. This makes clear that the problem is no longer just to simulate future payoffs, but to evaluate a conditional expectation at each exercise date and compare it with the exercise payoff.

This is the point where traditional Monte Carlo starts to become difficult, because the continuation term is the expectation conditional on the current state, so a direct simulation approach would require an inner simulation from each outer path and at each exercise date. If $N$ outer paths and $N$ inner paths are used, the work is already of order $N^2$ at one layer, and repeating this procedure over several exercise dates makes the computational cost grow very quickly. In \citep{Glasserman2004}, nested simulation is also used to illustrate how conditional expectations inside nonlinear payoffs create additional numerical difficulties, while in \citep{LongstaffSchwartz2001} the key object is the conditional expected payoff from continuation.

For that reason, the valuation of American and Bermudan options by Monte Carlo usually relies on approximations to the continuation value rather than on direct nested simulation. This is the point at which regression-based and related methods enter, and it motivates the next section.

\subsection{Longstaff-Schwartz and American Monte Carlo}

\subsubsection{Least-Squares Monte Carlo for American Option Pricing}

Regression-based approaches to early exercise already appear in \citep{Carriere1996,TsitsiklisVanRoy1997,TsitsiklisVanRoy2001,LongstaffSchwartz2001}, but in practice the Least-Squares Monte Carlo (LSM) framework of \citep{LongstaffSchwartz2001} became the standard reference, because it gave a simple and flexible way to treat American-style exercise within a simulation setting, without falling back on a full nested Monte Carlo scheme at every exercise date.

The main idea is to replace the exact continuation value, which would be too expensive to compute directly inside each simulated path, by an approximation obtained from cross-sectional regression on the simulated sample. In \citep{LongstaffSchwartz2001}, the conditional expected payoff from continuation is estimated from the realized future cash flows, allowing the exercise decision to be calculated in a backward way, starting from maturity, where the payoff is known, and moving step by step toward the initial date. At each exercise time, the model compares the immediate exercise payoff with the estimated continuation value, and this produces an approximate exercise rule together with an estimate of the option price.

This approach preserved the flexibility of Monte Carlo, while making it feasible in settings for pricing American/Bermudan options where the state space is large, the payoff is path dependent, or several factors drive the value of the contract. For that reason, least-squares Monte Carlo (LSM) became one of the main models for early-exercise options valuation \citep{LongstaffSchwartz2001}.

\subsubsection{American Monte Carlo for Exposure Estimation}

The same backward-looking logic can also be used beyond pricing at time zero. In \citep{CesariAquilinaCharpillon2009}, the counterparty exposure problem is approached as a pricing problem, and American Monte Carlo is introduced as an alternative to the classical forward Monte Carlo scheme when the product contains callability or other early-exercise features. Instead of producing only one price at time zero, the algorithm moves backward from maturity and generates values at the intermediate dates as well, which makes it possible to obtain the distribution of future values needed for exposure analysis.

This can be use for counterparty risk applications, because if the holder may exercise before maturity, future exposure is no longer determined only by the simulated market paths, because it also depends on whether the contract is still alive or has already been exercised. Once exercise takes place on a given path, the trade is no longer alive for that path, so its future exposure is zero from that point onward. The same estimates used to approximate continuation values for pricing can therefore be used to build future value profiles that are consistent with the exercise rule, and these profiles can then be taken into exposure measures and CVA calculations \citep{CesariAquilinaCharpillon2009,Gregory2015}.

In \citep{Gregory2015}, American Monte Carlo is also described as a practical optimisation approach used in xVA work, especially when the pricing overhead can be absorbed within the simulation through regressions on the options risk factors. Seen from the exposure side, the method follows the same logic as the original LSM, since the same step that approximates the exercise decision also gives the future path values on which exposure is based.

\subsubsection{Limitations of Classical Regression-Based Approaches}

The classical regression-based framework is flexible, but it also comes with a limitation that becomes more visible as the dimension of the problem grows. In \citep{HerreraKrachRuyssenTeichmann2023}, an issue is described in terms of basis functions, since the ordinary least-squares approximation requires a set of functions to be chosen in advance, and there is no single choice that works equally well across all products and all state spaces.

Additionally and more important for this article, in high-dimensional settings, the number of basis functions may grow exponentially with the dimension of the underlying process, which can make the classical approach increasingly difficult to use in practice. For this reason, other ways of approximating the continuation value have also been proposed in the literature, especially for problems where the dimensionality becomes harder to handle \citep{HerreraKrachRuyssenTeichmann2023}.

\subsection{Randomized Neural Networks for Optimal Stopping}

\subsubsection{Randomized Neural Networks}

In randomized neural networks, instead of adjusting all the weights of the network by back-propagation, part of the network is generated randomly and then kept fixed. In the feed-forward setting, this usually means that the hidden-layer weights and biases are randomly selected, while only the output weights are fitted afterwards. In \citep{CaoWangMingGao2018}, this type of model is described as a neural network with random weights, and the main motivation is to reduce the training complexity that appears when all parameters are tuned iteratively. A related result in \citep{HuangChenSiew2006} shows that single-hidden-layer feed-forward networks with randomly generated hidden nodes can still have universal approximation properties when the output weights are adjusted properly.

This idea is useful because the random hidden layer works as a feature map. The original state variables are transformed into a larger set of nonlinear signals, and the final model is built by fitting a linear combination of those signals. The nonlinearity therefore comes from the hidden layer, while the fitted part of the model remains much simpler than in a fully trained neural network. This is the same general reason why these methods are attractive in regression problems since they give a richer approximation space without turning the whole training procedure into a large non-convex optimization problem.

A similar separation appears in recurrent models under reservoir computing, where the reservoir is a recurrent system, often randomly generated and fixed, which processes an input sequence and produces internal states. The final output is then obtained by training a readout, often a linear one, on top of those states \citep{SchrauwenVerstraetenCampenhout2007,LukoseviciusJaeger2009}. This terminology is more common for recurrent networks than for feed-forward randomized networks, but the underlying idea is that the neurons structure is mostly fixed, and the trainable part is pushed to the final layer.

\subsubsection{Randomized Neural Networks for Continuation Value Approximation}

In the optimal stopping setting, the estimation problem that has to be approximated is still the option continuation value for each time step. The classical Longstaff-Schwartz method does this by regressing realized future cash flows on a set of basis functions chosen in advance while the  randomized neural network approach keeps the same backward induction structure, but changes the regression model used for the continuation value. Instead of working with a fixed polynomial basis, it uses a randomized neural network to generate random nonlinear features of the current state \citep{LongstaffSchwartz2001,HerreraKrachRuyssenTeichmann2023}.

This means that in the classical regression approach, the quality of the approximation depends heavily on the chosen basis functions, and adding more state variables has a direct effect on the model cost. In \citep{HerreraKrachRuyssenTeichmann2023}, this is one of the motivations for replacing the regression of the basis functions with randomized neural networks. The hidden-layer parameters are sampled randomly and kept fixed, while the last-layer parameters are estimated by least squares, since the continuation-value approximation is linear in those final parameters.

This keeps the estimation step close to ordinary least squares, but the representation of the state is no longer limited to the original variables or to a small polynomial basis. Additionally, the number of coefficients estimated in the final regression is not affected by the number of dimensions or state variables included in the input, meaning that more information can be added to the state representation without directly increasing the cost of the model, which is one reason why the method is presented as useful for high-dimensional stopping problems, where classical basis expansions can become difficult to manage \citep{HerreraKrachRuyssenTeichmann2023}.

The motivation of this approach is that it modifies the continuation-value approximation without changing the logic of the simulation and backward-induction process used by LSM. It therefore gives a direct way to compare the classical regression benchmark with a nonlinear random-feature approximation for pricing, the exposure profiles, and CVA estimates built from the resulting exercise policy.

\subsection{Black-Scholes and Heston models}

The numerical methods studied later in the article require a model for the evolution of the underlying risk factors. For this article two frameworks are considered, the first is the Black-Scholes-Merton model, which provides the classical lognormal approach, while the second is the Heston model, which provides a more realistic approach by implementing stochastic volatility dynamics.

\subsubsection{Black-Scholes-Merton model}

The first model was proposed by \citep{BlackScholes1973} and later extended by \citep{Merton1973} and has become one of the most important frameworks for options valuation. In \citep{BlackScholes1973}, the value of an option is derived from the absence of arbitrage, using the idea that a hedge position in the stock and the option should not allow for a risk-free profit, and in \citep{Merton1973}, the same idea is extended and used as part of a more general approach to option pricing. The importance of these papers is not only the final pricing formula, but also the option price dynamics that can be obtained from this model under a no-arbitrage assumption.

The model assumes that the stock price follows a continuous-time random walk with constant variance rate, so that future stock prices are lognormally distributed over a time \citep{BlackScholes1973}. In the risk-neutral representation, the expected growth rate of the stock is replaced by the risk-free rate, while volatility controls the dispersion of future prices. Assuming this, the option values can be expressed as discounted expectations under the risk-neutral measure, and Monte Carlo valuation can be done by simulating paths of the underlying asset, evaluating the discounted payoff on each path, and averaging across paths \citep{Glasserman2004}.

However, one limitation of this model is the constant volatility assumption, as this is not observed in real markets and ignores the implied volatility changes with the option strike and time to maturity \citep{Glasserman2004}. Despite this limitation, Black-Scholes model is still the most important for option pricing because it provides a simple methodology that explains the option price with the underlying price and allows for closed-form results.

\subsubsection{Heston Model}

The Heston model addresses this main limitation of the Black-Scholes-Merton framework by modeling a stochastic volatility insead of assuming it constant. In \citep{Heston1993}, the underlying price and its variance are modelled together, and this variance follows a mean-reverting square-root process. Therefore that volatility is no longer a fixed parameter assumed outside the model, but an stochastic state variable.

Consequently, the valuation problem is no longer driven only by the level of the underlying asset, since the current variance also affects the distribution used to price the option, while the model still leads to a closed-form solution for European option prices \citep{Heston1993}.

Another characteristic of the model is that it includes a correlation between stock returns and variance, so price movements and changes in variance are not treated as independent. In \citep{Heston1993}, this correlation explains the returns skewness and the strike-price biases in the Black-Scholes model, so the model changes not only the amount of uncertainty in future prices, but also how this uncertainty affects option values. In the numerical chapters, this gives a more complex scenario than Black-Scholes, since the simulated value depends on the stock path, the variance path and the relation between both sources of randomness.

\FloatBarrier
\clearpage

\section{Methodology} \label{methodology}

\subsection{Overall Simulation Framework}

In this article, the methodology used for the valuation and exposure estimation of American-style equity derivatives is based on a Monte Carlo framework, where the underlying variables are first simulated under a risk-neutral model, the exercise rule is then estimated by moving backwards through the exercise dates, and that estimated rule is later applied along simulated paths to obtain both the price and the future exposure profile. This follows the general simulation logic of Least-Squares Monte Carlo in \citep{LongstaffSchwartz2001}, where continuation values are approximated from simulated paths, while the later RLSM specification keeps the same stopping problem and replaces the classical regression basis with randomized neural-network features as in \citep{HerreraKrachRuyssenTeichmann2023}.

Following the finite exercise grid used for Bermudan approximations of American options in \citep{HerreraKrachRuyssenTeichmann2023}, the exercise dates are written as
\begin{equation}
0=t_0<t_1<\cdots<t_N=T,
\end{equation}
where $t_k$ denotes the $k$-th exercise date, $T$ is the maturity, and $N$ is the last exercise index, while in the numerical implementation the exposure is evaluated on this same grid, so exercise and exposure remain defined on the same set of dates.

At each exercise date, the intrinsic value is computed path by path. For a vanilla option written on one underlying price $S_t$, with strike $K$ and positive part $x^+=\max(x,0)$, the call and put payoffs are  \citep{Hull2003}
\begin{align}
g(S_t)^{\mathrm{call}}&=(S_t-K)^+,\\
g(S_t)^{\mathrm{put}}&=(K-S_t)^+.
\end{align}
For the multi-asset max-call contract used in the high-dimensional experiments, the payoff used in \citep{HerreraKrachRuyssenTeichmann2023} is 
\begin{equation}
g(S_t)=\left(\max_{1\leq j\leq d} S_t^{(j)}-K\right)^+,
\end{equation}
where $d$ is the number of underlying assets and $S_t^{(j)}$ is the value of asset $j$ at time $t$.

The exercise problem is treated in discrete time, and under the Snell-envelope recursion used in \citep{HerreraKrachRuyssenTeichmann2023}, the value at each exercise date is the larger of the immediate payoff and the value of continuation, which in the notation used here is
\begin{equation}
V_{t_k}(X_{t_k})=\max\big(g(X_{t_k}),C_{t_k}(X_{t_k})\big),
\end{equation}
where $V_{t_k}(X_{t_k})$ is the option value at $t_k$, $g(X_{t_k})$ is the immediate exercise payoff, and $C_{t_k}(X_{t_k})$ is the continuation value. Following the notation of \citep{HerreraKrachRuyssenTeichmann2023}, the continuation value is written as
\begin{equation}
C_{t_k}(x)=
\mathbb{E}^{\mathbb{Q}}
\left[
\alpha\,V_{t_{k+1}}(X_{t_{k+1}})
\mid X_{t_k}=x
\right],
\end{equation}
where $\mathbb{Q}$ is the risk-neutral measure and $\alpha$ denotes the one-step discount factor, which under the deterministic numeraire used here is
\begin{equation}
\alpha=\frac{N(t_k)}{N(t_{k+1})}=\exp\big(-r(t_{k+1}-t_k)\big).
\end{equation}

The continuation value in the previous equation is not computed directly, since the conditional expectation is unknown on the simulated paths, so both LSM and RLSM replace it with a regression estimate.

The workflow used in the implementation is the following.
\begin{enumerate}
\item \textbf{Contract and time grid.}
The product type, strike, maturity and exercise opportunities are fixed first, and the simulation grid is then chosen so that all relevant decision and exposure dates are included.

\item \textbf{Simulation of the state process.}
Under the chosen risk-neutral dynamics, Monte Carlo paths are generated for the underlying variables. In the Black-Scholes setting this means the spot process, while in the Heston setting both the spot and the variance processes are simulated.

\item \textbf{Pathwise payoff evaluation.}
At each exercise date and on each simulated path, the intrinsic payoff is computed, giving the immediate exercise values that are used later in the backward induction.

\item \textbf{Approximation of continuation values.}
Starting from the final exercise date and moving backwards, the continuation value is estimated as a conditional expectation of discounted future values. In the LSM case this is done by ordinary least-squares regression on a chosen basis, while in the RLSM case the same conditional expectation is approximated by the randomized neural network.

\item \textbf{Construction of the stopping rule.}
At each exercise date, the immediate payoff is compared with the estimated continuation value, and exercise takes place whenever the intrinsic value is higher than the value of continuation.

\item \textbf{Forward pricing and exposure evaluation.}
Once the stopping rule has been estimated, it is applied along simulated paths to compute the option price and the future exposure profile, and if exercise takes place on a path at time $t_k$, the contract is terminated on that path and the later exposures are set to zero.
\end{enumerate}

The pricing logic is the same in both approximation methods, but the Monte Carlo samples are handled differently. In the LSM model, one set of paths is used to estimate the regression functions and an independent set is used for pricing and exposure measurement, while for RLSM a single path set is generated and then split into training and evaluation subsets, following the same idea of separating fitted continuation values from the paths used to report the price.

After the stopping policy has been estimated, the price at time $0$ is obtained by moving forward along each evaluation path until the first exercise time, discounting the resulting cash flow and averaging over paths, as described in \citep{LongstaffSchwartz2001} and in the risk-neutral Monte Carlo pricing construction of \citep{Glasserman2004}. With $M$ evaluation paths and estimated stopping time $\tau^{(i)}$ on path $i$, the estimator used here is
\begin{equation}
\widehat{V}_0
=
\frac{1}{M}\sum_{i=1}^{M}
\frac{H_{\tau^{(i)}}^{(i)}}{N(\tau^{(i)})},
\end{equation}
where $H_{\tau^{(i)}}^{(i)}$ is the realised payoff on path $i$ at its stopping time and $N(\tau^{(i)})$ discounts the payoff back to the initial date.

From the institution's point of view, credit exposure is the positive value of the transaction \citep{Gregory2010CCR}. For path $i$ and future date $t_k$, this is represented as
\begin{equation}
E^{(i)}(t_k)=\max\left(V_{t_k}^{(i)},0\right),
\end{equation}
where $V_{t_k}^{(i)}$ is the pathwise option value at that date. If the option has already been exercised on that path, the implementation sets the later pathwise value, and therefore the exposure, equal to zero.

At a fixed future time, expected exposure is just the expected value of exposure \citep{Gregory2010CCR}. In the Monte Carlo sample this is estimated by
\begin{equation}
\widehat{EE}(t_k)=\frac{1}{M}\sum_{i=1}^{M}E^{(i)}(t_k),
\end{equation}
where the same $M$ evaluation paths are used as in the pricing step. Potential Future Exposure is a quantile of the future value distribution in \citep{CesariAquilinaCharpillon2009}; applied to the exposure distribution at time $t_k$, it is written as
\begin{equation}
PFE_q(t_k)
=
\inf\left\{x\in\mathbb{R}:\mathbb{P}(E(t_k)\le x)\ge q\right\},
\end{equation}
where $q\in(0,1)$ is the selected confidence level and $E(t_k)$ denotes the random exposure at that date.

Since the same product, path generator, exercise grid and aggregation rules are kept in both methods, the comparison between LSM and RLSM is mainly a comparison of how the continuation value is approximated.

\clearpage
\subsection{Path Generation Models} \label{path gen}

The pricing and exposure calculations are run on simulated paths of the risk factors, using two path models throughout the numerical experiments. In the Black-Scholes case, the state variable is the spot price with constant volatility, whereas in the Heston case each path contains both the spot price and its stochastic variance. Both models are simulated under the risk-neutral measure $\mathbb{Q}$ and use the deterministic numeraire introduced in the previous subsection, with the single-asset versions used for vanilla American call and put options and the multi-asset versions used for the max-call experiments.

\subsubsection{Black-Scholes model}

The single-asset Black-Scholes case is simulated under the risk-neutral measure, and since no dividend yield is included in this implementation, the spot process follows the geometric Brownian motion dynamics described in \citep{Glasserman2004},
\begin{equation}
dS_t=rS_t\,dt+\sigma S_t\,dW_t,
\end{equation}
where $S_t$ is the stock price at time $t$, $r$ is the constant risk-free rate, $\sigma>0$ is the volatility and $W_t$ is a Brownian motion under $\mathbb{Q}$. The path is generated from the exact lognormal transition over each interval $[t_k,t_{k+1}]$ \citep{Glasserman2004},
\begin{equation}
S_{t_{k+1}}
=
S_{t_k}
\exp\left(
\left(r-\frac{1}{2}\sigma^2\right)\Delta t_k
+
\sigma\sqrt{\Delta t_k}\,Z_{k+1}
\right),
\end{equation}
where $\Delta t_k=t_{k+1}-t_k$ and $Z_{k+1}$ is a standard normal random variable. Because the transition is exact for this model, the Black-Scholes paths are not generated with an Euler discretisation.

For the max-call experiments, the multi-asset Black-Scholes path model applies this lognormal transition component by component, so with $d$ assets, dividend yield $q_j$ and volatility $\sigma_j$ for asset $j$, and using the risk-neutral drift $r-q_j$ as in \citep{Glasserman2004}, the update is
\begin{equation}
S_{t_{k+1}}^{(j)}
=
S_{t_k}^{(j)}
\exp\left(
\left(r-q_j-\frac{1}{2}\sigma_j^2\right)\Delta t_k
+
\sigma_j\sqrt{\Delta t_k}\,Z_{k+1}^{(j)}
\right),
\end{equation}
for $j=1,\dots,d$, where $Z_{k+1}^{(1)},\dots,Z_{k+1}^{(d)}$ are independent standard normal shocks. 
\begin{algorithm}
\caption{Black-Scholes path generation}
\begin{algorithmic}[1]
\Require Simulation grid $0=t_0<t_1<\cdots<t_N=T$; model parameters
$S_0$, $r$, $\sigma$; in the multi-asset case, the vectors
$(q_j,\sigma_j)_{j=1}^d$
\Ensure Simulated path $\{S_{t_k}\}_{k=0}^{N}$ or
$\{S_{t_k}^{(j)}\}_{k=0,\dots,N}^{j=1,\dots,d}$
\State Set the initial state equal to the spot value(s) at time $t_0$
\For{$k=0,\dots,N-1$}
    \State Compute $\Delta t=t_{k+1}-t_k$
    \State Generate independent standard normal shocks
    $Z$ (single asset) or $Z_1,\dots,Z_d$ (multi-asset)
    \State Update the asset value(s) using the exact lognormal transition
    \State Store the simulated value(s) at time $t_{k+1}$
\EndFor
\end{algorithmic}
\end{algorithm}

\subsubsection{Heston model}

For Heston, the simulations use the stochastic-volatility setup from the randomized optimal stopping experiments of \citep{HerreraKrachRuyssenTeichmann2023}, based on the model of \citep{Heston1993}, and the single-asset dynamics are
\begin{align}
dS_t &= (r-q)S_t\,dt+\sqrt{v_t}\,S_t\,dW_t^S,\\
dv_t &= -\kappa(v_t-\theta)\,dt+\sigma\sqrt{v_t}\,dW_t^v,\\
d\langle W^S,W^v\rangle_t &= \rho\,dt.
\end{align}
Here $S_t$ is the spot price, $v_t$ is its variance, $q$ is the dividend yield, $\theta$ is the long-run variance level, $\kappa$ is the mean-reversion speed, $\sigma$ is the volatility of variance and $\rho$ is the correlation between the Brownian shocks.

The Euler step uses the Heston setup in \citep{HerreraKrachRuyssenTeichmann2023}, and the initial variance is set equal to the long-run variance level, $v_{t_0}=\theta$. Let $v_{t_k}^{+}=\max(v_{t_k},0)$, let $Z_{k+1}^S$ and $Z_{k+1}^{\perp}$ be independent standard normal shocks, and define the variance shock as
\begin{equation}
Z_{k+1}^v=\rho Z_{k+1}^S+\sqrt{1-\rho^2}\,Z_{k+1}^{\perp}.
\end{equation}
For a time step $\Delta t_k=t_{k+1}-t_k$, the variance is updated first and the spot equation then uses the newly computed variance \citep{HerreraKrachRuyssenTeichmann2023}.
\begin{align}
v_{t_{k+1}}
&=
v_{t_k}
-
\kappa\big(v_{t_k}^+-\theta\big)\Delta t_k
+
\sigma\sqrt{v_{t_k}^+}\sqrt{\Delta t_k}\,Z_{k+1}^v,\\
S_{t_{k+1}}
&=
S_{t_k}
+
(r-q)S_{t_k}\Delta t_k
+
\sqrt{v_{t_{k+1}}^+}\,S_{t_k}\sqrt{\Delta t_k}\,Z_{k+1}^S.
\end{align}
The positive part $v^+=\max(v,0)$ is used to ensure obtaining positive volatilities and to avoid taking the square root of negative values, and then the spot update uses the updated positive variance $v_{t_{k+1}}^+$. In the multi-asset Heston experiments, the same process is applied component by component in order to generate the max-call underlying assets risk factors.

\begin{algorithm}
\caption{Heston path generation}
\begin{algorithmic}[1]
\Require Simulation grid $0=t_0<t_1<\cdots<t_N=T$; model parameters
$S_0$, $r$, $q$, $\theta$, $\kappa$, $\sigma$, and $\rho$; in the multi-asset case,
their asset-specific vector counterparts
\Ensure Simulated path of spot and variance states
\State Initialize the spot process at $S_0$
\State Initialize the variance process at $v_0=\theta$
\For{$k=0,\dots,N-1$}
    \State Compute $\Delta t=t_{k+1}-t_k$
    \State Generate two independent standard normal shocks
    \State Construct the correlated variance shock using $\rho$
    \State Update the variance using the positive part of the previous variance
    \State Update the spot using the positive part of the updated variance
    \State Store the simulated spot and variance values at time $t_{k+1}$
\EndFor
\end{algorithmic}
\end{algorithm}

\clearpage
\subsection{Least-Squares Monte Carlo (LSM)} \label{lsm}

\subsubsection{Introduction and role of LSM}

This Least-Squares Monte Carlo model uses the simulated risk factors paths introduced in Section \ref{path gen} and works backward over the exercise dates in order to approximate continuation values and construct an exercise policy \citep{LongstaffSchwartz2001}. In this article, that same policy is then applied to an independent path sample to obtain both the price and the future value profile from which exposure measures are computed. The state vector at time $t_k$, denoted by $X_{t_k}$, contains the spot under Black-Scholes dynamics and the spot together with the variance under Heston dynamics, so the regression step always acts on the market variables generated by the chosen path model.

In discrete time, if $t_k$ is an exercise date, $h_{t_k}(x)$ denotes the immediate exercise payoff at state $x$ and $C_{t_k}(x)$ the continuation value, then the continuation value can be written as \citep{Glasserman2004}
\begin{equation}
C_{t_k}(x)=\mathbb{E}\left[V_{t_{k+1}}(X_{t_{k+1}})\mid X_{t_k}=x\right].
\end{equation}
while the corresponding option value is
\begin{equation}
V_{t_k}(x)=\max\left(h_{t_k}(x),C_{t_k}(x)\right).
\end{equation}
and, for the products considered here, this conditional expectation is not available in closed form, so LSM replaces it with a regression estimate computed from simulated paths, after which the fitted stopping rule can be used to calculate the pathwise future values and then aggregated to build the exposure profiles used in counterparty-risk \citep{CesariAquilinaCharpillon2009,Gregory2015}.

\subsubsection{Two-stage Monte Carlo design}

The implementation uses two independent Monte Carlo samples, a pre-simulation sample for the continuation-value regressions and a main-simulation sample for the evaluation of the fitted exercise policy. The reason for this split is that the stopping rule is estimated on one path set and then evaluated on another one, so the reported values are not computed on the same sample used to fit the regressions, which helps reduce look-ahead effects and the risk of an in-sample upward bias in the valuation results \citep{LongstaffSchwartz2001}.

The simulated paths are generated on a grid that contains both the exercise dates and the exposure dates, so if $\mathcal{T}^{\mathrm{ex}}=\{t_0,\dots,t_N\}$ denotes the exercise grid and $\mathcal{T}^{\mathrm{exp}}=\{u_0,\dots,u_L\}$ the exposure grid, the simulation timeline is defined by
\begin{equation}
\mathcal{T}^{\mathrm{sim}}=\mathrm{sort}\!\left(\mathcal{T}^{\mathrm{ex}}\cup\mathcal{T}^{\mathrm{exp}}\right).
\end{equation}
Using the same timeline keeps the stopping policy and the future-value evaluation on the same grid, so exercise and exposure quantities are computed consistently across dates.

\clearpage
\subsubsection{Regression approximation of continuation values}

At each exercise date, the conditional expectation defining the continuation value is approximated by a linear combination of basis functions of the current state, so if $\phi_{k,1},\dots,\phi_{k,p_k}$ denote the basis functions used at date $t_k$, the regression representation is
\begin{equation}
\mathbb{E}\!\left[V_{t_{k+1}}(X_{t_{k+1}})\mid X_{t_k}=x\right]=\sum_{m=1}^{p_k}\beta_{k,m}\phi_{k,m}(x),
\end{equation}
which can be written more compactly as
\begin{equation}
C_{t_k}(x)=\beta_k^\top\phi_k(x),
\end{equation}
where $\beta_k=(\beta_{k,1},\dots,\beta_{k,p_k})^\top$ collects the regression coefficients and $\phi_k(x)=\\(\phi_{k,1}(x),\dots,\phi_{k,p_k}(x))^\top$ the basis functions evaluated at the current state \citep{Glasserman2004}.

The coefficient vector associated with this approximation satisfies
\begin{equation}
\beta_k=
\left(
\mathbb{E}\!\left[\phi_k(X_{t_k})\phi_k(X_{t_k})^\top\right]
\right)^{-1}
\mathbb{E}\!\left[\phi_k(X_{t_k})V_{t_{k+1}}(X_{t_{k+1}})\right],
\end{equation}
where the expectations are taken under the joint distribution of $(X_{t_k},X_{t_{k+1}})$ generated by the underlying Markov state process \citep{Glasserman2004}.

For the pre-simulation sample $\{(X_{t_k}^{(j)},X_{t_{k+1}}^{(j)})\}_{j=1}^{M_{\mathrm{pre}}}$, these expectations are replaced by their sample counterparts, so the least-squares estimator is
\begin{equation}
\widehat{\beta}_k=\widehat{B}_{\phi,k}^{-1}\widehat{B}_{\phi V,k},
\end{equation}
with
\begin{equation}
\widehat{B}_{\phi,k}
=
\frac{1}{M_{\mathrm{pre}}}
\sum_{j=1}^{M_{\mathrm{pre}}}
\phi_k(X_{t_k}^{(j)})\phi_k(X_{t_k}^{(j)})^\top,
\qquad
\widehat{B}_{\phi V,k}
=
\frac{1}{M_{\mathrm{pre}}}
\sum_{j=1}^{M_{\mathrm{pre}}}
\phi_k(X_{t_k}^{(j)})\widehat{V}_{t_{k+1}}(X_{t_{k+1}}^{(j)}),
\end{equation}
and the resulting continuation estimate is
\begin{equation}
\widehat{C}_{t_k}(x)=\widehat{\beta}_k^\top\phi_k(x).
\end{equation}
The specific choice of basis functions depends on the underlying path model and is described in the next subsection \citep{Glasserman2004}.

\subsubsection{Basis functions used in the regression}

The basis functions implemented here are the same power basis functions used in the LSM model of \citep{HerreraKrachRuyssenTeichmann2023}, and their precise specification depends on the underlying path model.
Polynomial bases are a standard choice in regression-based American Monte Carlo methods, and products interaction terms are added for multi-dimensional products \citep{Glasserman2004,HerreraKrachRuyssenTeichmann2023}.

In the single-asset Black-Scholes case, where the state contains only the spot value, the basis is
\begin{equation}
\Phi^{\mathrm{pow}}(S_t)=\left[1,S_t,S_t^2,\dots,S_t^{d_{\mathrm{reg}}}\right]^\top.
\end{equation}
For multi-asset Black-Scholes, the spot variables are first normalized by the strike,
\begin{equation}
\widetilde{S}_t^{(j)}=\frac{S_t^{(j)}}{K},\qquad j=1,\dots,d,
\end{equation}
and the interaction terms between the underlying spots are included when $d_{\mathrm{reg}}\geq 2$.

Under Heston dynamics, the regression state contains both spot and variance, and these variables are normalized as
\begin{equation}
\widetilde{S}_t=\frac{S_t}{K},
\qquad
\widetilde{v}_t=\frac{v_t}{v_{\mathrm{ref}}},
\end{equation}
where $v_{\mathrm{ref}}$ is a strictly positive reference variance. In the single-asset case, the basis is
\begin{equation}
\Phi^{\mathrm{pow}}(\widetilde{S}_t,\widetilde{v}_t)=\left[1,\widetilde{S}_t,\widetilde{S}_t^2,\dots,\widetilde{S}_t^{d_{\mathrm{reg}}},\widetilde{v}_t,\widetilde{v}_t^2,\dots,\widetilde{v}_t^{d_{\mathrm{reg}}}\right]^\top.
\end{equation}
For the multi-asset Heston setting, the basis is extended component by component and includes powers of the normalized spot variables, powers of the normalized variance variables and the interactions terms for the spots and the variances states when $d_{\mathrm{reg}}\geq 2$.

\subsubsection{Backward induction and stopping rule}

The backward recursion starts from the final exercise date and then moves step by step toward the initial date, so at maturity there is no continuation value left and the terminal condition is
\begin{equation}
\widehat{V}_{t_N}(X_{t_N}^{(i)})=H_{t_N}(X_{t_N}^{(i)}).
\end{equation}
For earlier exercise dates, the estimated continuation values define the exercise decision path by path, and the recursion is written as
\begin{equation}
\widehat{V}_{t_k}(X_{t_k}^{(i)})=
\begin{cases}
H_{t_k}(X_{t_k}^{(i)}), & \text{if } H_{t_k}(X_{t_k}^{(i)})\geq \widehat{C}_{t_k}(X_{t_k}^{(i)}),\\
\widehat{V}_{t_{k+1}}(X_{t_{k+1}}^{(i)}), & \text{if } H_{t_k}(X_{t_k}^{(i)})< \widehat{C}_{t_k}(X_{t_k}^{(i)}),
\end{cases}
\end{equation}
which means that exercise takes place as soon as the immediate payoff is at least as large as the estimated continuation value, while otherwise the contract is continued to the next date \citep{Glasserman2004,LongstaffSchwartz2001}.

Since the products considered here are single-right contracts, exercise permanently terminates the cash flows, and for the later exposure calculations it is needed to introduce an alive indicator $A_{t_k}^{(i)}\in\{0,1\}$, where $A_{t_k}^{(i)}=1$ means that path $i$ is still active at date $t_k$ and $A_{t_k}^{(i)}=0$ means that exercise has already taken place. Its update is
\begin{equation}
A_{t_{k+1}}^{(i)}=
A_{t_k}^{(i)}\,
\mathbf{1}_{\{H_{t_k}(X_{t_k}^{(i)})<\widehat{C}_{t_k}(X_{t_k}^{(i)})\}},
\qquad
A_{t_0}^{(i)}=1,
\end{equation}
so the indicator remains equal to one only if the contract is alive at time $t_k$ and continuation is chosen at that date, while if exercise occurs at $t_k$ the indicator becomes zero from the next step onward and all later continuation values and cash flows on that path are set to zero \citep{LongstaffSchwartz2001}.

\subsubsection{Forward pricing and exposure extraction}

Once the backward step has calculated the exercise policy, this policy is applied to the out of sample on the main-simulation paths, and each path contributes with the discounted exercise payoff at its estimated stopping time, so the realized discounted payoff is written as
\begin{equation}
\Pi^{(i)}=
\frac{H_{\widehat{\tau}^{(i)}}\!\left(X_{\widehat{\tau}^{(i)}}^{(i)}\right)}{N\!\left(\widehat{\tau}^{(i)}\right)},
\end{equation}
where $\widehat{\tau}^{(i)}$ denotes the estimated stopping time on path $i$. Averaging these discounted payoffs gives the price at time $0$ price estimator
\begin{equation}
\widehat{V}_0=
\frac{1}{M_{\mathrm{main}}}
\sum_{i=1}^{M_{\mathrm{main}}}
\Pi^{(i)}.
\end{equation}
This is the same forward evaluation step used after the exercise decisions have been determined along each path, where one moves to the first stopping time, discounts the resulting exercise cash flow back to time $0$, and then averages across paths, while in the regression-based formulation the continuation estimates define the exercise policy that is followed on a second path sample \citep{LongstaffSchwartz2001,Glasserman2004}.

For exposure calculations, the relevant object at an exposure date $u_\ell$ is the future value conditional on the option still being alive, which in counterparty-risk terminology known as the mark-to-market at that future date, so the pathwise future value is defined as
\begin{equation}
FV^{(i)}(u_\ell)=
A_{u_\ell}^{(i)}
\frac{\widehat{V}_{u_\ell}\!\left(X_{u_\ell}^{(i)}\right)}{N(u_\ell)}.
\end{equation}
This uses the alive indicator to ensures that only contracts that remain active contribute to the future value, while contracts that have already been exercised no longer generate counterparty exposure, and the continuation estimate is therefore used only on paths for which the option is still alive at date $u_\ell$ \citep{LongstaffSchwartz2001,CesariAquilinaCharpillon2009}.

The pathwise exposure is then obtained from the positive part of the future value,
\begin{equation}
E^{(i)}(u_\ell)=\max\left(FV^{(i)}(u_\ell),0\right),
\end{equation}
and from these pathwise exposures it is immediate to compute profiles such as expected exposure by averaging across paths and potential future exposure by taking the appropriate quantile of the future-value or exposure distribution \citep{Gregory2015,CesariAquilinaCharpillon2009}.

\clearpage
\subsubsection{LSM Algorithm}

\begin{algorithm}
\caption{Least-Squares Monte Carlo with out-of-sample pricing and exposure}
\begin{algorithmic}[1]
\Require Exercise dates $\mathcal{T}^{\mathrm{ex}}$, exposure dates $\mathcal{T}^{\mathrm{exp}}$, contract payoff $h$, state generator, numeraire $N(\cdot)$, regression basis $\Phi$, pre-simulation size $M_{\mathrm{pre}}$, main-simulation size $M_{\mathrm{main}}$
\Ensure Price at time $0$, pathwise exposure profile, and PFE profile
\State Build the common simulation grid $\mathcal{T}^{\mathrm{sim}}=\mathrm{sort}(\mathcal{T}^{\mathrm{ex}}\cup\mathcal{T}^{\mathrm{exp}})$
\State Generate $M_{\mathrm{pre}}$ pre-simulation paths on $\mathcal{T}^{\mathrm{sim}}$
\State Initialize the terminal value at the last exercise date with the immediate payoff
\For{regression dates $t_k$ in backward order}
    \State Construct the future normalized cashflow targets on the pre-simulation sample
    \State Build the design matrix from the simulated state at $t_k$
    \State Fit the least-squares regression for the continuation value
    \State Store the fitted model for the corresponding exercise or exposure date
\EndFor
\State Generate $M_{\mathrm{main}}$ independent main-simulation paths on $\mathcal{T}^{\mathrm{sim}}$
\State Initialize the alive state and the vector of discounted cash flows
\For{exposure dates $u_\ell$ in forward order}
    \State Apply all exercise opportunities up to and including $u_\ell$
    \State Update the alive state and record the discounted exercise cash flow when exercise occurs
    \State Evaluate the fitted continuation value at $u_\ell$ under the current alive state
    \State Convert the future value at $u_\ell$ into pathwise exposure
\EndFor
\State Compute the price at time $0$ as the sample mean of realized discounted cash flows
\State Compute the EE profile from the positive part of the pathwise exposures
\State Compute the PFE profile as the empirical pathwise quantile at each date
\end{algorithmic}
\end{algorithm}

\clearpage
\subsection{Randomized Least-Squares Monte Carlo (RLSM)} \label{rlsm}

\subsubsection{Introduction and role of RLSM}

The objective of the implementation used in this article is to extend the randomized neural-network model for optimal stopping proposed by \citep{HerreraKrachRuyssenTeichmann2023} so that it can also be used to estimate exposure profiles, and for that purpose some modifications are introduced with respect to the original model, although the general structure of the method remains the same as in \citep{HerreraKrachRuyssenTeichmann2023}. In particular, the problem is still the backward optimal stopping problem for a Bermudan approximation of the American contract, so the economic logic is unchanged and the difference with Section \ref{lsm} lies in how the continuation value is approximated.

Using the notation of \citep{HerreraKrachRuyssenTeichmann2023}, the option value at exercise date $n$ is written as
\begin{equation}
U_n=\max\big(g(X_n),c_n(X_n)\big),
\end{equation}
where $g(X_n)$ denotes the immediate exercise payoff and the continuation value is
\begin{equation}
c_n(x)=\mathbb{E}\big[\alpha U_{n+1}\mid X_n=x\big].
\end{equation}
This is the same stopping problem considered in the original randomized least-squares formulation, and the adaptations introduced in this article affect the state input, the readout fit, and the reconstruction of future values for exposure calculations, while the backward structure of the method is kept unchanged \citep{HerreraKrachRuyssenTeichmann2023}.

\subsubsection{Randomized continuation-value approximation}

In the original RLSM formulation, the continuation value is approximated by a randomized neural network in which the hidden layer is sampled once and kept fixed, while only the parameters of the last layer are fitted at each exercise date. Using the notation of \citep{HerreraKrachRuyssenTeichmann2023}, the random feature map is defined by
\begin{equation}
\phi:\mathbb{R}^{d}\to\mathbb{R}^{K},\qquad x\mapsto\phi(x)=\big(\sigma(Ax+b)^{\top},1\big)^{\top},
\end{equation}
where $\sigma$ is the activation function, $A$ and $b$ are the random hidden parameters, and the final entry equal to $1$ accounts for the intercept term.

The continuation value at time $n$ is then approximated by
\begin{equation}
c_{\theta_n}(x):=\theta_n^{\top}\phi(x)=A_n^{\top}\sigma(Ax+b)+b_n,
\end{equation}
where $\theta_n=((A_n)^{\top},b_n)^{\top}$ contains the parameters of the linear readout. The hidden parameters are therefore not optimized, and the approximation problem is reduced to fitting the last layer only \citep{HerreraKrachRuyssenTeichmann2023}.

The readout is fitted by least squares and, at time $n$, this is done by minimizing the following loss function

\begin{equation}
\psi_n(\theta_n):=
\sum_{i=1}^{m}
\left(
c_{\theta_n}(x_n^i)-\alpha p_{n+1}^i
\right)^2,
\end{equation}
whose minimizer is given in closed form by
\begin{equation}
\theta_n=
\alpha
\left(
\sum_{i=1}^{m}
\phi(x_n^i)\phi^{\top}(x_n^i)
\right)^{-1}
\cdot
\left(
\sum_{i=1}^{m}
\phi(x_n^i)p_{n+1}^i
\right).
\end{equation}
This is the main structural difference with the polynomial regression used in Section \ref{lsm}, because the nonlinear approximation is generated by the randomized hidden map, while the fitted problem at each exercise date is still a linear least-squares solve in the readout parameters only. As a result, the number of fitted coefficients in RLSM does not depend on the dimensionality of the underlying problem, whereas in the LSM model the number of regression coefficients directly grows with the number of dimensions. Additionally, the randomized hidden representation can  capture nonlinearities more flexibly than the fixed polynomial basis \citep{HerreraKrachRuyssenTeichmann2023}.

\subsubsection{State representation}

In \citep{HerreraKrachRuyssenTeichmann2023}, the hidden map is applied to the current state, and the implementation used in this article keeps that idea but enriches the input before the random projection. This is done because, in the exposure calculations, a richer input tends to improve convergence than it does for the price, where the differences are usually much smaller. Under Black-Scholes dynamics, the input contains the current spot $S_n$, the current payoff $g(S_n)$, and powers of both quantities up to degree three, whereas under Heston dynamics it also contains the current variance $v_n$ together with same-asset spot-variance interaction terms, again with powers up to degree three. 

The current payoff $g(S_n)$ is include as an input because it helps to identify the local exercise region, and \citep{HerreraKrachRuyssenTeichmann2023} reports that RLSM usually works slightly better when the payoff is included. The time to maturity, $\tau_n=T-t_n$, may also be added to the input, although its effect is less consistent, since in some configurations it improves the fit while in others the results are worse.

\subsubsection{Train-test split and hidden-state construction}

RLSM uses one single Monte Carlo sample of $2m$ paths, the first $50\%$ of the sampled paths $\{1,2,\dots,m\}$, as training data to fit the weights and bias, and then using the remaining $50\%$ $\{m+1,\dots,2m\}$, as evaluation data for the option price and exposure estimation. The reason for this train-test split is that the continuation value is a conditional expectation and should not depend on future values from the same sample on which it is evaluated, so separating the training and evaluation paths helps avoid that dependence and reduces the tendency to overfit the realized future cash flows \citep{HerreraKrachRuyssenTeichmann2023}.

Once the paths are simulated and state input has been created, it is passed through the randomized hidden layer for all simulated paths before the backward recursion starts, so in the backward step the model works directly with those hidden features at each date and it does not need to repeat the same process for each iteration.

\subsubsection{Regularized readout}

While \citep{HerreraKrachRuyssenTeichmann2023} considers the use of regularization when the number of hidden neurons are large, during the results it is said that a better approach is directly reducing the hidden layer. However, the estimation future exposure is more sensitive to noise, and the larger number of inputs increases the risk of overfitting, therefore the regularization step becomes necessary. To control the effects of this regularization, other changes implemented consist in the standardization of the hidden features, the non-regularization of the bias term, and the normalization of the ridge penalty by the training sample size.

To explain these implementations in a compact way, let $Z_n\in\mathbb{R}^{m\times H}$ be the matrix of hidden features on the training paths at exercise date $n$, and let $Y_n\in\mathbb{R}^{m}$ be the vector of continuation targets. The hidden features and targets are standardized by their means,
\begin{equation}
\widetilde{Y}_n=Y_n-\bar{Y}_n,
\qquad
\widetilde{Z}_{n,j}=\frac{Z_{n,j}-\bar{Z}_{n,j}}{\sigma_{n,j}},
\end{equation}
where $\bar{Y}_n$ is the sample mean of the targets, $\bar{Z}_n$ collects the column means of $Z_n$, and and $\sigma_{n,j}$ is its sample standard deviation. This standardization is necessary because with the implemented inputs, the hidden features end up on different scales and the effect of the regularization can vanish.

With this notation, the penalized regression can be written as
\begin{equation}
\widehat{w}_n=
\arg\min_{w\in\mathbb{R}^{H}}
\left\{
\frac{1}{m}\sum_{i=1}^{m}\left(\widetilde{Y}_{n,i}-(\widetilde{Z}_n w)_i\right)^2+\lambda\sum_{j=1}^{H}w_j^2
\right\},
\end{equation}
where $w=(w_1,\dots,w_H)^\top$ is the vector of linear coefficients fitted on the hidden features, $\widetilde{Z}_n$ and $\widetilde{Y}_n$, are the standardized hidden-feature matrix and continuation targets, $\lambda\geq0$ is the ridge parameter, and $1/m$ is the normalization by the training sample size. After this fit, the coefficients are rescaled back to the original units, while the intercept is reconstructed separately and is therefore not penalized. In this way, the three main changes introduced in this article are the standardization of the hidden features, the normalization of the ridge term by the training sample size, and the separate treatment of the bias term.

\subsubsection{Backward recursion}

The next step after the continuation approximation is fitted for each date, is the backward recursion \citep{HerreraKrachRuyssenTeichmann2023}. At maturity, the option value on each simulated path is equal to the payoff,
\begin{equation}
p_N^i=g(x_N^i), \qquad i\in\{1,2,\dots,2m\},
\end{equation}
where $p_n^i$ denotes the price approximation at date $n$ on path $i$, and $x_n^i$ is the simulated state on that path and date.

For each earlier date $n\in\{N-1,N-2,\dots,0\}$, the recursion compares the immediate exercise value with the estimated continuation value and writes
\begin{equation}
p_n^i :=
\begin{cases}
g(x_n^i), & \text{if } g(x_n^i)\geq c_{\theta_n}(x_n^i),\\
\alpha p_{n+1}^i, & \text{otherwise},
\end{cases}
\qquad i\in\{1,2,\dots,2m\},
\end{equation}
so if the payoff is larger than the continuation estimate the path is exercised at date $n$, while otherwise the value is the discounted next-step price. Therefore, the stopping decision is already part of the recursion process, and no additional exercise rule is needed \citep{HerreraKrachRuyssenTeichmann2023}.

\subsubsection{Price and exposure evaluation}

After the backward recursion, the price at time $0$ is computed on the evaluation half of the sample, following the same final averaging step used in \citep{HerreraKrachRuyssenTeichmann2023},
\begin{equation}
p_0=\max\left(g(x_0),\frac{1}{m}\sum_{i=m+1}^{2m}\alpha p_1^i\right),
\end{equation}
where $p_0$ is the price at time $0$ approximation, and the average is taken over the evaluation paths only.

For exposure calculations, the future value at exposure date $u_\ell$ is written with the same notation as in Section \ref{lsm},
\begin{equation}
FV^{(i)}(u_\ell)=
A_{u_\ell}^{(i)}
\frac{\widehat{V}_{u_\ell}\!\left(X_{u_\ell}^{(i)}\right)}{N(u_\ell)},
\end{equation}
where $A_{u_\ell}^{(i)}$ is the alive indicator on path $i$, $\widehat{V}_{u_\ell}(X_{u_\ell}^{(i)})$ is the estimated option value at the exposure date, and $N(u_\ell)$ is the numeraire. The same alive indicator of Section \ref{lsm} is used so that the only contracts that remain active are considered for the future value, while paths that have already been exercised no longer generate counterparty exposure.

The pathwise exposure is then obtained from the positive part of the future value,
\begin{equation}
E^{(i)}(u_\ell)=\max\left(FV^{(i)}(u_\ell),0\right),
\end{equation}
and from these pathwise values the exposure profiles used later in the CVA framework are obtained in the same way as in Section \ref{lsm}.

\subsubsection{RLSM Algorithm}

\begin{algorithm}
\caption{RLSM training, pricing and exposure evaluation}
\begin{algorithmic}[1]
\Require Simulation grid, exercise dates, exposure dates, payoff $g$, hidden size $H$, activation $\sigma$, ridge parameter $\lambda$, total path count $2m$
\Ensure Price at time $0$, pathwise exposures, EE profile, PFE profile
\State Simulate one Monte Carlo sample on the full time grid
\State Build the state input at all dates and for all simulated paths
\State Randomly initialize the hidden map $\Psi(x)=\sigma(Wx+b)$
\State Evaluate the hidden representation for all paths and all time steps
\State Split the paths into the training set $\{1,\dots,m\}$ and the evaluation set $\{m+1,\dots,2m\}$
\For{exercise dates in backward order}
    \State Build the continuation targets on the training set
    \State Fit the linear coefficients on the hidden features, with the regularized fit described above
    \State Update the pathwise values using the backward recursion
\EndFor
\State Compute the price at time $0$ on the evaluation set
\State Initialize the alive indicator on the exposure paths
\For{exposure dates in forward order}
    \State Apply the exercise decisions up to the current date
    \State Reconstruct the future value on alive paths
    \State Set the future value to zero on exercised paths
    \State Record the discounted pathwise exposure
\EndFor
\State Compute the EE profile from the pathwise exposures
\State Compute the PFE profile from the empirical quantiles of the exposure distribution
\end{algorithmic}
\end{algorithm}

\clearpage
\subsection{CVA Framework}

\subsubsection{Framework assumptions}

The CVA framework used in this article takes the simulated exposure profiles produced by the pricing models and combines them with a deterministic description of counterparty default, so the market component and the credit component remain separated throughout the calculation. This follows the usual decomposition in which counterparty risk is introduced as an adjustment to the risk-free valuation, while the exposure side is generated independently by the underlying pricing engine \citep{Gregory2015}.

The setting is unilateral, in the sense that only the default of the counterparty is taken into account, while Debit Valuation Adjustment (DVA), collateral, margin period of risk, funding effects, and wrong-way risk are left outside the present framework. This choice is taken because the purpose of the article is to keep the credit component fixed and simple, so that the effect of the exposure, and the difference between LSM and RLSM, can be observed more clearly in the final CVA results.

\subsubsection{Portfolio aggregation and netting} \label{portfolio agg}

For CVA, we need the net exposure of the portfolio because in the presence of a netting agreement, positive and negative trade values offset each other, and therfore the final exposure is based on the netted portfolio value.
If $V_i(t)$ denotes the value of trade $i$ at time $t$, the contract-level exposure is
\begin{equation}
E_i(t)=\max\{V_i(t),0\},
\end{equation}
so the exposure of one trade is just its positive value at time $t$ \citep{ZhuPykhtin2007}. When several trades belong to the same netting agreement, denoted here by $NA$, the exposure becomes
\begin{equation}
E_{\mathrm{net}}(t)=\max\left\{\sum_{i\in NA}V_i(t),0\right\},
\end{equation}
so positive and negative values are offset first, and the resulting net positive amount is used for the credit exposure \citep{ZhuPykhtin2007,Gregory2010CCR}.

On the Monte Carlo exposure grid, the same rule is applied path by path, so portfolio value is
\begin{equation}
V_{\mathrm{port}}^{(i)}(t_k)=\sum_{j=1}^{J} w_j\,V_j^{(i)}(t_k),
\end{equation}
where $w_j$ is the weight of trade $j$ in the portfolio and $i$ denotes the Monte Carlo path,
and the corresponding portfolio exposure is
\begin{equation}
E_{\mathrm{port}}^{(i)}(t_k)=\max\left(V_{\mathrm{port}}^{(i)}(t_k),0\right),
\end{equation}
and after calculating the average of the paths we can get the expected exposure of the portfolio,
\begin{equation}
EE_{\mathrm{port}}(t_k)=\frac{1}{M}\sum_{i=1}^{M}E_{\mathrm{port}}^{(i)}(t_k).
\end{equation}
where $M$ is the number of simulated paths \citep{Gregory2010CCR}. This is the last step needed to calculate the market risk component of the CVA equation.

\subsubsection{Deterministic hazard-rate model and default probabilities}

The credit side of the framework is represented through a deterministic hazard-rate curve, which keeps the treatment of default separate from the exposure simulation. Following the standard reduced-form setting, the counterparty default time $\tau$ is described through a deterministic hazard rate $\gamma(t)$ and cumulative hazard function
\begin{equation}
\Gamma(t)=\int_0^T \gamma(t)\,du.
\end{equation}
Under this assumption, the survival probability up to time $t$ is
\begin{equation}
\mathbb{Q}(\tau>t)=\exp(-\Gamma(t)),
\end{equation}
and the default probability over an interval $(s,t]$ is
\begin{equation}
\mathbb{Q}(s<\tau\le t)=\exp(-\Gamma(s))-\exp(-\Gamma(t))
\end{equation}
\citep{Brigo2005}. These relations are the ones used later when the simulated exposure profile is combined with default probabilities.

In the implementation, the hazard-rate curve is specified on a set of market maturities $0<T_1<\cdots<T_m$ and is taken piecewise constant between consecutive nodes, which is a standard practical choice when implied default probabilities are bootstrapped from CDS data \citep{Gregory2010CCR,Brigo2005}. Writing the hazard levels as $\gamma_1,\dots,\gamma_m$, this means
\begin{equation}
\gamma(t)=
\begin{cases}
\gamma_1, & 0\le t\le T_1,\\
\gamma_2, & T_1<t\le T_2,\\
\vdots\\
\gamma_m, & T_{m-1}<t\le T_m.
\end{cases}
\end{equation}
The cumulative hazard at a given time is obtained by adding the hazard contributions of the intervals up to that time, so the integral above is reduced to a sum over the relevant segments. Given the exposure grid $t_0<t_1<\cdots<t_k$, the marginal default probability $PD(t_{k-1},t_k)$ used in each interval is therefore
\begin{equation}
PD(t_{k-1},t_k)=\mathbb{Q}(t_{k-1}<\tau\le t_k)=\exp(-\Gamma(t_{k-1}))-\exp(-\Gamma(t_k)).
\end{equation}
which provides the default probabilities used in the discrete CVA approximation later on \citep{ZhuPykhtin2007}.

\subsubsection{Unilateral CVA estimator}

Unilateral CVA is the expected discounted loss generated by the default of the counterparty, so once the exposure profile and the default probabilities have been specified, both parts can be combined in the usual reduced-form way. In the notation of \citep{ZhuPykhtin2007}, this is written as
\begin{equation}
\mathrm{CVA}=(1-R)\int_0^T EE^*(t)\,dPD(0,t),
\end{equation}
where $R$ is the recovery rate, $PD(0,t)$ is the cumulative default probability up to time $t$, and $EE^*(t)$ denotes the discounted expected exposure, defined by
\begin{equation}
EE^*(t)=\mathbb{E}^Q\left[\frac{B_0}{B_t}E(t)\right],
\end{equation}
where $\mathbb{E}^Q$ is expectation under the risk-neutral measure, $E(t)$ is the exposure at time $t$, and $B_t$ is the value of the numeraire at time $t$ \citep{ZhuPykhtin2007}.

On the discrete exposure grid used in the Monte Carlo implementation, the integral is replaced by a sum over the exposure dates, which gives
\begin{equation}
\mathrm{CVA}=(1-R)\sum_{k=1}^{n} EE^*(t_k)\,PD(t_{k-1},t_k).
\end{equation}
Adapting it to the the portfolio expected exposure from Section \ref{portfolio agg}, we get the portfolio CVA by
\begin{equation}
\mathrm{CVA_{port}}=(1-R)\sum_{k=1}^{n} EE_{\mathrm{port}}(t_k)\,PD(t_{k-1},t_k).
\end{equation}
This is the final form used later in the case study, where the exposure model determines the terms $EE_{\mathrm{port}}(t_k)$ and the hazard-rate curve determines the terms $PD(t_{k-1},t_k)$ \citep{ZhuPykhtin2007,Brigo2005}.

\subsubsection{Monte Carlo error estimation}

Since we use Monte Carlo averaging for the CVA estimation, the error can be calculated from the dispersion of the pathwise CVA contributions. If $\mathrm{CVA}^{(i)}$ denotes the contribution of path $i$, the estimator is
\begin{equation}
\mathrm{CVA}=\frac{1}{M}\sum_{i=1}^{M}\mathrm{CVA}^{(i)},
\end{equation}
and the corresponding Monte Carlo standard error is
\begin{equation}
SE(\mathrm{CVA})=
\frac{1}{\sqrt{M}}
\left(
\frac{1}{M-1}\sum_{i=1}^{M}\left(\mathrm{CVA}^{(i)}-\mathrm{CVA}\right)^2
\right)^{1/2}.
\end{equation}
where $M$ is the number of simulated paths \citep{Glasserman2004}. This Monte Carlo error is given with the CVA results in the numerical results of Section \ref{cva case study}.

\subsubsection{CVA Algorithm}

\begin{algorithm}[H]
\caption{CVA framework}
\begin{algorithmic}[1]
\Require Pathwise trade values on the exposure grid, portfolio weights, exposure dates, recovery rate $R$, hazard-rate curve
\Ensure CVA estimate and Monte Carlo standard error
\State Aggregate trade values path by path to obtain portfolio values on the exposure grid
\State Apply the positive part to obtain pathwise portfolio exposures
\State Average across paths to obtain the portfolio expected exposure profile
\State Compute the interval default probabilities $PD(t_{k-1},t_k)$ from the hazard-rate curve
\State Combine the exposure profile and the interval default probabilities to compute the pathwise CVA contributions
\State Average the pathwise contributions to obtain the CVA estimate
\State Estimate the Monte Carlo standard error from the sample dispersion of the pathwise CVA contributions
\end{algorithmic}
\end{algorithm}

\clearpage
\section{Experiment Design} \label{experiment design}

During this section we will describe the configuration of the experiments performed in Section \ref{experiment results} and \ref{cva case study}, where we will compare Least-Squares Monte Carlo (LSM) and Randomized Least-Squares Monte Carlo (RLSM) under the same configuration in order to answer the research questions of the article. The choice of experiment configuration is based on the one used by \citep{HerreraKrachRuyssenTeichmann2023}, because this allow us to obtain comparable results and assess whether the same conclusions hold. The main purpose of the experiments is to check whether the RLSM specification adopted in this article converges to the same
price and exposure outputs as the LSM benchmark for low-dimensional and high-dimensional products.

\subsection{Products}

Durign the experiment section, two different products are covered. The first one is the traditional \emph{vanilla American options} with a single underlying equity, this gives the low-dimensional benchmark and it is used in Section \ref{vanilla options}.

The second is formed a \emph{high-dimensional American max-call options} with the underlying vector $S_t=(S_t^{(1)},\dots,S_t^{(d)})$, and a payoff at time $t$
\begin{equation}
H_t=
\max\!\left(
\max_{1\le j\le d} S_t^{(j)}-K,\,0
\right).
\end{equation}
The max-call option is a standard example in the literature on high-dimensional American option pricing, it appears for instance in \citep{LongstaffSchwartz2001,Glasserman2004,LapeyreLelong2021,BeckerCheriditoJentzen2019}, and it is also the main product used in \citep{HerreraKrachRuyssenTeichmann2023}. This makes it a natural choice here, since it lets us keep the experiments close to Herrera et al. while moving from the one-dimensional vanilla case to a setting where the dimension of the state space is the main difficulty. For this reason, it is the product used here in the high-dimensional experiments.

\subsection{Model Setup}

All experiments are carried out under either Black-Scholes dynamics or Heston stochastic-volatility dynamics. The baseline parameter configurations are chosen according to the ones used in the benchmark setup used in \citep{HerreraKrachRuyssenTeichmann2023}, so that the numerical results remain directly comparable.

\subsubsection{Black-Scholes configuration}

Under the Black-Scholes model, the underlying price process follows
\begin{equation}
dS_t=(r-q)S_t\,dt+\sigma S_t\,dW_t,
\end{equation}
with constant volatility $\sigma$, constant short rate $r$, and dividend yield $q$ \citep{BlackScholes1973,Merton1973}. The baseline specification used in the experiments is
\begin{equation}
S_0=100,
\qquad
K=100,
\qquad
T=1,
\qquad
r=0,
\qquad
\sigma=0.2,
\qquad
q=0.
\end{equation}
In the multi-asset max-call experiments, the same spot, volatility and dividend yield are assigned component by component to each underlying, and the choice $q=0$ makes the comparison with the corresponding European price easier to interpret.

\subsubsection{Heston configuration}

Under the Heston model, the spot and variance processes satisfy
\begin{equation}
dS_t=(r-q)S_t\,dt+\sqrt{v_t}\,S_t\,dW_t^S,
\end{equation}
\begin{equation}
dv_t=
\kappa(\theta-v_t)\,dt
+
\sigma\sqrt{v_t}\,dW_t^v,
\end{equation}
with
\begin{equation}
d\langle W^S,W^v\rangle_t=\rho\,dt,
\end{equation}
which is the classical stochastic-volatility specification of \citep{Heston1993}. The baseline parameter configuration is
\begin{equation}
S_0=100,
\qquad
K=100,
\qquad
T=1,
\qquad
r=0,
\qquad
q=0,
\end{equation}
\begin{equation}
\theta=0.01,
\qquad
\kappa=2.0,
\qquad
\sigma=0.2,
\qquad
\rho=-0.3.
\end{equation}
The initial variance is set equal to the long-run variance level,
\begin{equation}
v_0=\theta=0.01.
\end{equation}

As in the Black-Scholes case, the multi-asset max-call experiments apply the same parameter values component by component across assets. Again, these parameters define a common benchmark and keep the setup close to \citep{HerreraKrachRuyssenTeichmann2023}, and the choice $q=0$ here as well help for the comparison with the price of the equivalent European option.

\subsection{Monte Carlo Configuration}

The numerical results are obtained from repeated Monte Carlo experiments under a common baseline path budget. In the baseline configuration, both methods use $m=10000$ paths, which keeps the setup close to \citep{HerreraKrachRuyssenTeichmann2023} and keeps the simulation manageable.

The use of this simulated paths the methodology of each algorithm, while in LSM the paths are used through the pre-simulation and main-simulation structure described in Section \ref{lsm}, in RLSM the same budget is used together with the train/evaluation split described in Section \ref{rlsm}. 

All reported results are obtained from repeated Monte Carlo runs with different random seeds, so that the comparison is not driven by one single simulation scenario.

\subsection{Baseline Model Parameters}

The baseline model parameters are chosen to remain close to \citep{HerreraKrachRuyssenTeichmann2023}, and the effect of hidden dimension and regularization is left for the sensitivity analysis, so here we only fix the values used in the main experiments.

For RLSM, the hidden dimension $H$ changes with the dimensionality of the contract, in one to five dimensions the baseline uses $H=25$, in the 10-dimensional case it uses $H=50$, and in the higher-dimensional cases it uses $H=100$, while the ridge parameter $\lambda$ is taken in the range from $0$ to $10^{-5}$ depending on the product and the dimension. The baseline configuration also uses:
\begin{itemize}
\item input factor equal to $1.0$,
\item leaky-ReLU activation function,
\item train/evaluation split equal to $2$,
\end{itemize}
Apart from the regularized readout refinements introduced in this article, the rest of the architectural choices follows the benchmark implementation used in \citep{HerreraKrachRuyssenTeichmann2023}.

For LSM, the baseline regression uses polynomial degree $3$, as this is the standard choice for LSM and provides stable results on exposure profiles .

\subsection{Time Discretization}

In the baseline experiments, the maturity is fixed at $T=1$, and the time grid is taken to be uniform with $N=20$ dates over $[0,T]$, including the initial date and maturity, so that
\begin{equation}
0=t_0<t_1<\dots<t_{N-1}=T.
\end{equation}
The same grid is used for the exercise dates, for the simulation of the underlying paths, and for the exposure observation dates, which keeps stopping, valuation and exposure on the same set of times.

\subsection{Reference Values}

The reference used in the experiments depends on the product class. In the one-dimensional vanilla American cases, a large-sample LSM run is still feasible, so this is the reference price used for the comparison.

In the high-dimensional American max-call options, because of computational limitations a high-precision American references are no longer feasible, so the European option price is used as in \citep{HerreraKrachRuyssenTeichmann2023}. Under the zero-dividend setting used here, this gives a simple reference of scale for the call-type products.

These references are enough for the purpose of this article, since our objective is to check whether RLSM converges to the same values as LSM and to remain consistent with high-dimensional \citep{HerreraKrachRuyssenTeichmann2023} for the high-dimensional scenario.

\subsection{Evaluation Metrics}

The comparison between methods is based on price differences, exposure-curve differences, dispersion across Monte Carlo runs, and computational time.

\subsubsection{Price accuracy}

Let $\widehat{P}^{(r)}$ denote the price estimate produced by a given method on Monte Carlo replication $r$, and let $P^{\mathrm{ref}}$ denote the benchmark price used for the comparison. Price differences are summarized through the mean absolute error
\begin{equation}
\mathrm{MAE}_{P}=
\frac{1}{R}\sum_{r=1}^{R}
\left|\widehat{P}^{(r)}-P^{\mathrm{ref}}\right|,
\end{equation}
where $R$ is the number of independent Monte Carlo replications.

\subsubsection{Exposure-curve accuracy}

For the exposure curves, let $\widehat{EE}^{(r)}(t_\ell)$ denote the expected exposure curve obtained on replication $r$, evaluated at the monitoring dates $t_0,\dots,t_L$, and let $EE^{\mathrm{ref}}(t_\ell)$ denote the benchmark curve. The RMSE is defined by
\begin{equation}
\mathrm{RMSE}_{EE}^{(r)}=
\left(
\frac{1}{L+1}\sum_{\ell=0}^{L}
\left(
\widehat{EE}^{(r)}(t_\ell)-EE^{\mathrm{ref}}(t_\ell)
\right)^2
\right)^{1/2}.
\end{equation}
The average curve error is then summarized by
\begin{equation}
\overline{\mathrm{RMSE}}_{EE}=
\frac{1}{R}\sum_{r=1}^{R}\mathrm{RMSE}_{EE}^{(r)}.
\end{equation}

\subsubsection{Dispersion measures}

Since all reported quantities are obtained by Monte Carlo simulation, dispersion across independent runs is reported together with the mean values whenever possible. For scalar outputs this is measured through the sample standard deviation \citep{Glasserman2004}
\begin{equation}
s_X=
\left(
\frac{1}{R-1}\sum_{r=1}^{R}
\left(X^{(r)}-\overline{X}\right)^2
\right)^{1/2},
\end{equation}
and through the associated standard error
\begin{equation}
\mathrm{SE}(X)=\frac{s_X}{\sqrt{R}}.
\end{equation}
This is the same standard Monte Carlo error measure used in \citep{HerreraKrachRuyssenTeichmann2023}.

\subsection{Computational Time}

Computational efficiency is assessed through the runtime, in the low- and high-dimensional results this runtime corresponds only to the valuation and exposure estimation, while path generation is not taken into the timed section, so that the results mainly reflect the differences in the models and allows for fair comparison, as in \citep{HerreraKrachRuyssenTeichmann2023} .

\clearpage
\section{Experiment Results} \label{experiment results}
\subsection{Vanilla American Options} \label{vanilla options}

In this section we show the results for vanilla American call options as the low-dimensional scenario of the article. This is not the scenario where RLSM is expected to have an advantage, because the stopping problem is relatively simple, the continuation value can already be approximated well with a small polynomial basis, and the number of inputs remains limited, therefore here LSM have a good performance while it remains efficient, while the complexity of RLSM may add noise.

\subsubsection{Black-Scholes Results}

The results for Black-Scholes are shown in Table~\ref{tab:vanilla_bs_results}. Both methods give almost exactly the same price, LSM gives $7.7764$ with a standard error $0.1130$, while RLSM gives $7.7766$ with a standard error $0.1486$. Both are very close to the reference value of $7.7934$, so from a pricing point of view the two methods are converging to the same results.

\begin{table}[H]
\centering
\begin{tabular}{lcccc}
\toprule
Model & Price$_0$ & Reference & EE$(0)$ & Time (s) \\
\midrule
LSM  & 7.7764 (0.1130) & 7.7934 & 7.7974 & 0.2430 \\
RLSM & 7.7766 (0.1486) & 7.7934 & 7.7999 & 0.4457 \\
\bottomrule
\end{tabular}
\caption[Vanilla American call results under Black-Scholes dynamics.]{Vanilla American call results under Black-Scholes dynamics. Standard errors are reported in parentheses. Source: Own elaboration.}
\label{tab:vanilla_bs_results}
\end{table}

\begin{figure}[H]
\centering
\includegraphics[width=0.70\textwidth]{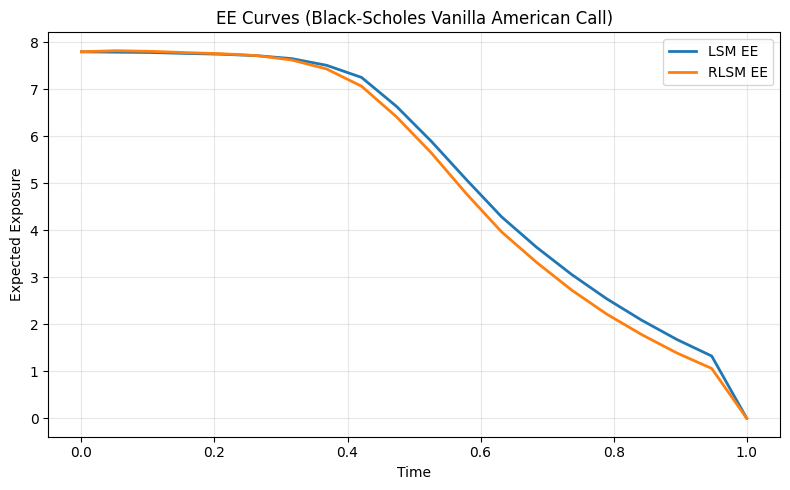}
\caption[Expected exposure curves for the vanilla American call under Black-Scholes dynamics.]{Expected exposure curves for the vanilla American call under Black-Scholes dynamics. Source: Own elaboration.}
\label{fig:vanilla_bs_ee}
\end{figure}

Regarding the Expected Exposure (EE) curve plot in Figure~\ref{fig:vanilla_bs_ee} we can see similar results. The
two curves remain close throughout the time, although the gap in exposure is
a bit more visible than the gap in price. Close to maturity the RLSM curve drops a
little faster, which points to slightly earlier exercise along some paths. In this scenario RLSM converges, but it does not bring a real advantage as LSM is also
faster, with an average runtime of $0.2430$ seconds against $0.4457$ seconds.
That is what we expected for the one-dimensional scenario where the
regression problem remains small and the classical basis already does the job well.

\subsubsection{Heston Results}

For the Heston model, the results are slightly different
Table~\ref{tab:vanilla_heston_results}. LSM produces a price of $2.5294$ with
standard error $0.0325$, almost identical to the reference value of $2.5314$, while
RLSM, gives $2.4762$ with a larger standard error of
$0.0471$, underestimating the price comparing to LSM.

\begin{table}[H]
\centering
\begin{tabular}{lcccc}
\toprule
Model & Price$_0$ & Reference & EE$(0)$ & Time (s) \\
\midrule
LSM  & 2.5294 (0.0325) & 2.5314 & 2.5313 & 0.3897 \\
RLSM & 2.4762 (0.0471) & 2.5314 & 2.4815 & 0.6961 \\
\bottomrule
\end{tabular}
\caption[Vanilla American call results under Heston dynamics.]{Vanilla American call results under Heston dynamics. Standard errors are reported in parentheses. Source: Own elaboration.}
\label{tab:vanilla_heston_results}
\end{table}

\begin{figure}[H]
\centering
\includegraphics[width=0.7\textwidth]{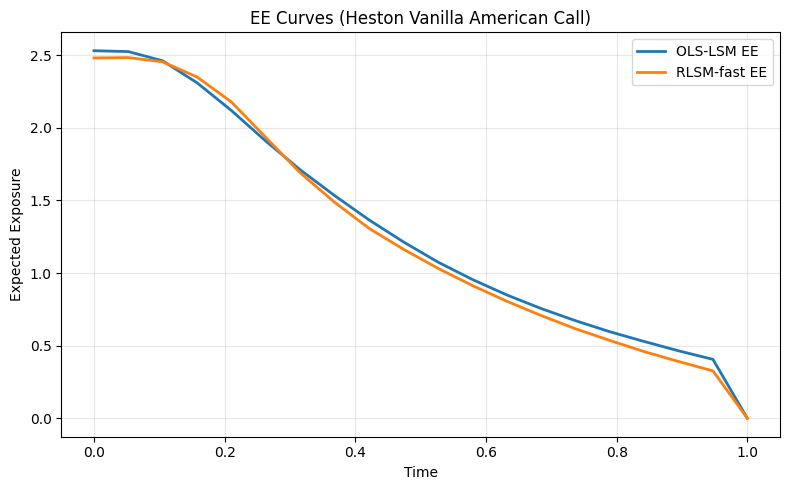}
\caption[Expected exposure curves for the vanilla American call under Heston dynamics.]{Expected exposure curves for the vanilla American call under Heston dynamics. Source: Own elaboration.}
\label{fig:vanilla_heston_ee}
\end{figure}

For the EE curve in Figure~\ref{fig:vanilla_heston_ee}, the lower RLSM price is also reflected in a lower initial exposure level.
However, the shape of the curve converges better to LSM than in the Black-Scholes case, and the runtime
results points in the same direction as for Black-Scholes, LSM takes on average $0.3897$ seconds,
whereas RLSM takes $0.6961$ seconds, so once again the randomized approach more expensive in low dimension. 

\subsubsection{PFE Results}

\begin{figure}[H]
     \centering
     \begin{subfigure}[b]{0.48\textwidth}
         \centering
         \includegraphics[width=\textwidth]{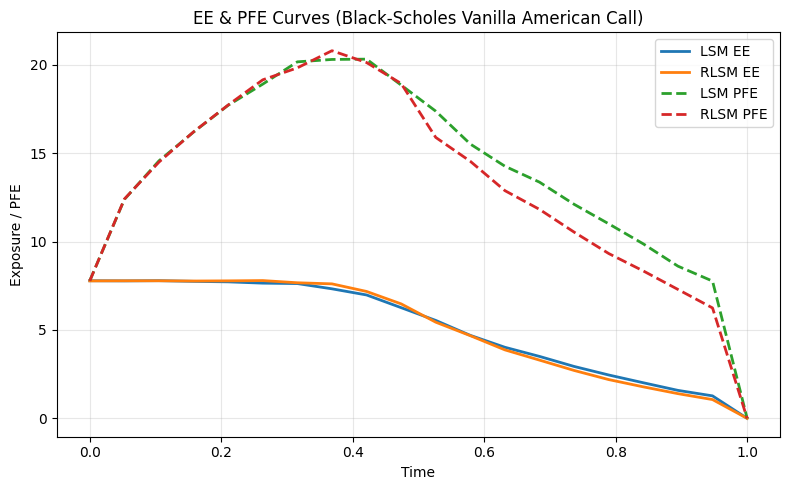}
         \caption{Black-Scholes dynamics.}
         \label{fig:vanilla_bs}
     \end{subfigure}
     \hfill 
     \begin{subfigure}[b]{0.48\textwidth}
         \centering
         \includegraphics[width=\textwidth]{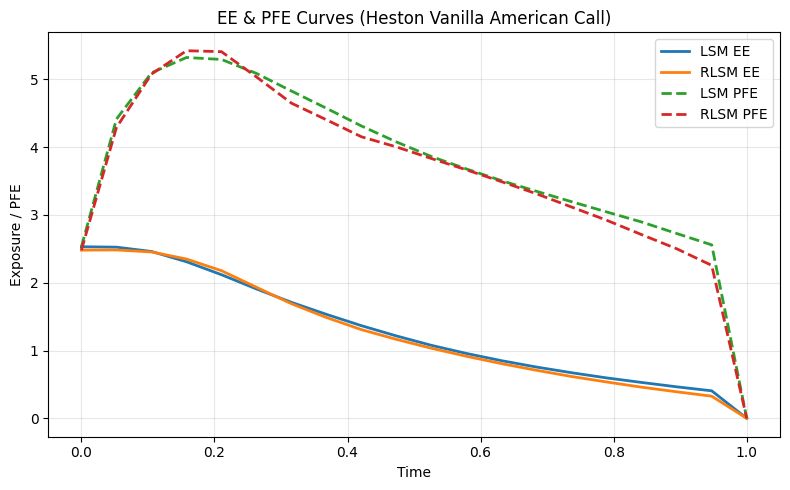}
         \caption{Heston dynamics.}
         \label{fig:vanilla_heston}
     \end{subfigure}
     
     \caption[EE and PFE comparison for the vanilla American call.]{EE and PFE comparison for the vanilla American call. Source: Own elaboration.}
     \label{fig:pfe_bs_heston}
\end{figure}

Moving to the Potential Future Exposure (PFE) results, in Figures~\ref{fig:pfe_bs_heston} the EE and PFE curves are shown. Here the same conclusions can be obtained as for EE curves, RLSM still shows convergence for PFE, however the differences become more visible in PFE than in EE because PFE is a tail measure (in this case it is $95\%$ quantile), so even relatively small discrepancies in the continuation value approximation or in the stopping policy become more pronounced once one moves from the mean profile to the upper tail of the exposure distribution.

This vanilla american option experiment points in the same direction under both
market models, RLSM converges to the results of LSM, but it is not the most
appropriate choice for this kind of product. In low dimension American options still have very linear relations and for these scenarios the classical LSM
regression is already accurate, stable and cheap, while the random features and non linear functions of RLSM introduces noise that affects negatively to the estimation. Additionally, for one dimension, LSM is able to converge to the reference price and EE for this number of paths, however, this sample size becomes insufficient as the dimensionality grows, a point further analyzed in Sections \ref{high dimension} and \ref{path conv}. Therefore, the strengths of RLSM appear later, in
problems that are more nonlinear and much higher-dimensionality.

\clearpage
\subsection{High-Dimensional Max-Call Options} \label{high dimension}

Here the scenario where the advantage of RLSM and the limitations of LSM should be more visible. Following the structure used in \citep{HerreraKrachRuyssenTeichmann2023},
we consider American max-call options under Black-Scholes and Heston dynamics,
with dimensions ranging from $d=5$ to $d=50$.

For LSM, the regression problem becomes progressively harder as the state
space expands, since the polynomial basis needed to retain a reasonable fit
quickly becomes more costly and more difficult to condition numerically, while for
RLSM, the number of fitted coefficients is tied
to the hidden layer size and does not grow directly with the number of state
variables. Therefore, during this experiment we tried to prove this points. 

\subsubsection{Black-Scholes Results}

The Black-Scholes results are reported in
Table~\ref{tab:maxcall_bs_results}. A first positive sign for RLSM is that the
pricing gap between both methods is now small across all dimensions, for
$d=5$, the LSM price is $24.3441$ while RLSM gives $24.2202$, for $d=10$, the
results are $33.3940$ and $33.3412$, for $d=25$, $44.9478$ and $44.8581$, and
for $d=50$, $52.7926$ and $53.0506$. So, despite the increase in dimensionality,
the two methods remain close in price. As already observed mentioned in \citep{HerreraKrachRuyssenTeichmann2023},
RLSM often tends to produce slightly lower prices, although the effect here is
small and does not alter the overall picture \citep{HerreraKrachRuyssenTeichmann2023}.

\begin{table}[H]
\centering
\begin{tabular}{lccccc}
\toprule
Model & $d$ & Price$_0$ & Euro MC & EE$(0)$ & Time (s) \\
\midrule
LSM  & 5  & 24.3441 (0.1097) & 24.9604 & 24.3768 & 1.10 \\
RLSM & 5  & 24.2202 (0.1853) & 24.9604 & 24.2969 & 0.40 \\
LSM  & 10 & 33.3940 (0.1196) & 34.3058 & 33.5566 & 2.16 \\
RLSM & 10 & 33.3412 (0.2206) & 34.3058 & 33.5106 & 0.79 \\
LSM  & 25 & 44.9478 (0.1408) & 45.9492 & 45.5024 & 12.13 \\
RLSM & 25 & 44.8581 (0.2439) & 45.9492 & 45.1066 & 1.31 \\
LSM  & 50 & 52.7926 (0.1468) & 54.3279 & 54.5384 & 108.77 \\
RLSM & 50 & 53.0506 (0.1411) & 54.3279 & 53.1473 & 1.93 \\
\bottomrule
\end{tabular}
\caption[American max-call results under Black-Scholes dynamics.]{American max-call results under Black-Scholes dynamics. Standard errors are reported in parentheses. Source: Own elaboration.}
\label{tab:maxcall_bs_results}
\end{table}

\begin{figure}[H]
\centering

\begin{subfigure}{0.45\textwidth}
    \centering
    \includegraphics[width=\linewidth]{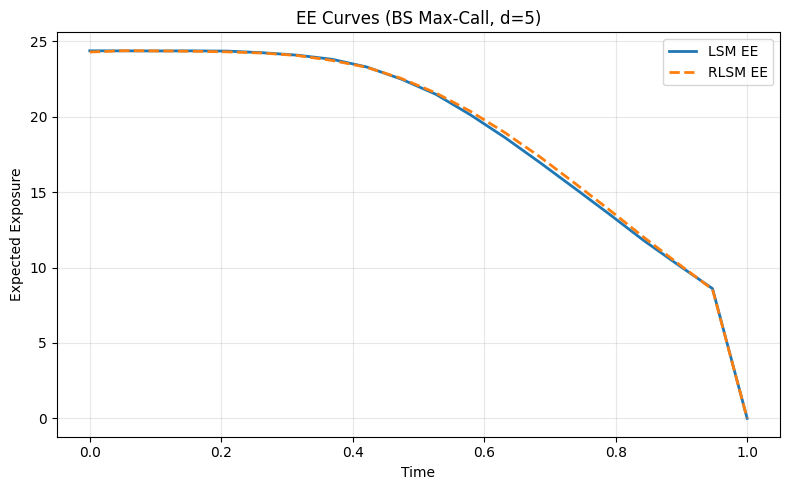}
    \caption{d = 5}
\end{subfigure}
\hfill
\begin{subfigure}{0.45\textwidth}
    \centering
    \includegraphics[width=\linewidth]{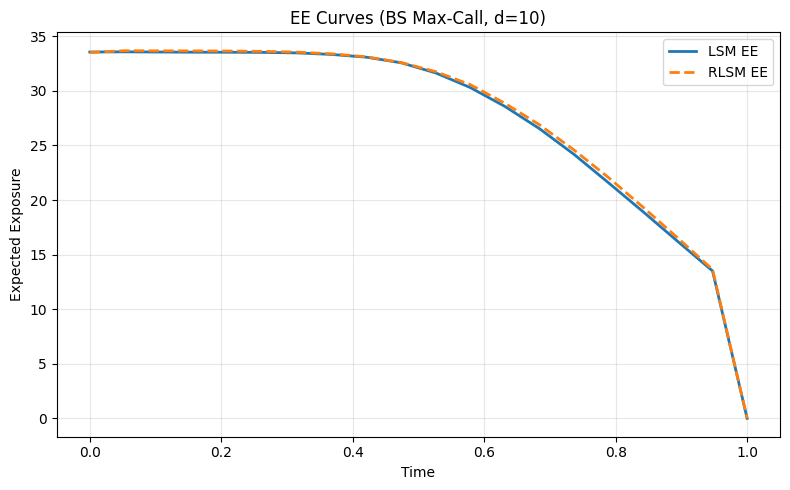}
    \caption{d = 10}
\end{subfigure}

\vspace{0.3cm}

\begin{subfigure}{0.45\textwidth}
    \centering
    \includegraphics[width=\linewidth]{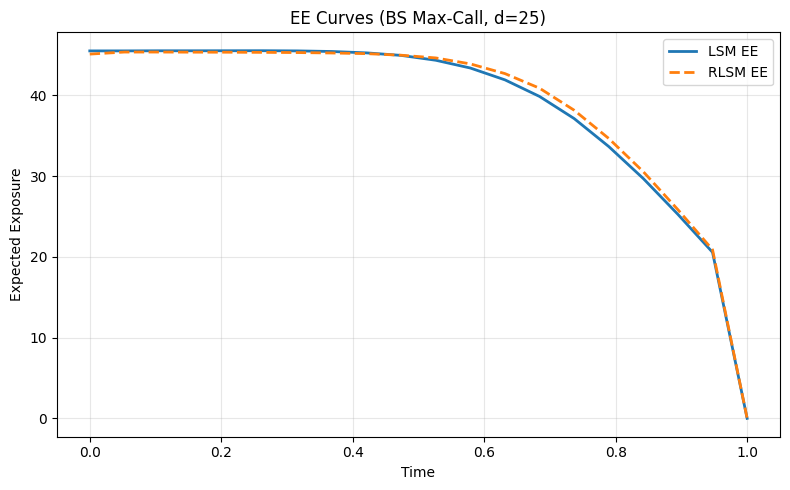}
    \caption{d = 25}
\end{subfigure}
\hfill
\begin{subfigure}{0.45\textwidth}
    \centering
    \includegraphics[width=\linewidth]{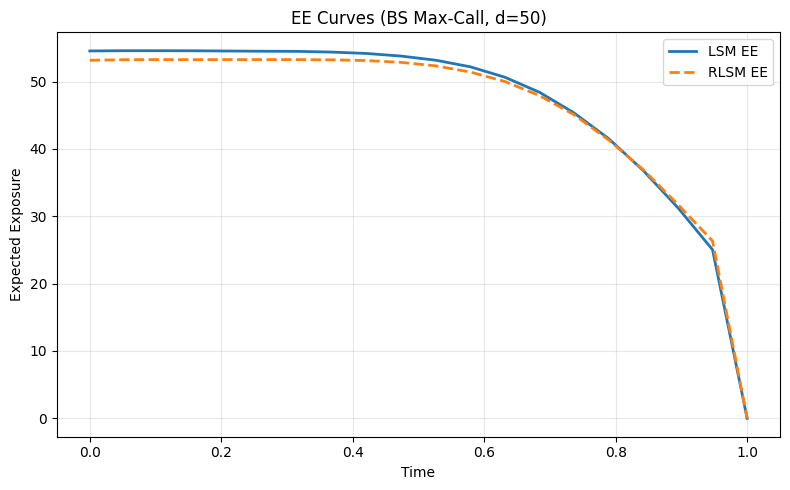}
    \caption{d = 50}
\end{subfigure}

\caption[EE curves for American max-call options under Black-Scholes dynamics across dimensions.]{EE curves for American max-call options under Black-Scholes dynamics across dimensions. Source: Own elaboration.}
\label{fig:maxcall_bs_ee}

\end{figure}

The EE curves in Figure~\ref{fig:maxcall_bs_ee} reinforce the same conclusion. For all four
dimensions, the RLSM exposure profiles are very close to the ones of LSM. Therefore, once we move into a nonlinear multi-asset setting, the
randomized approximation fully converges to the LSM
benchmark across the full exposure horizon.

At the same time, the runtime advantage becomes more clear, even at
$d=5$, RLSM is already faster than LSM, and the gap widens quickly as the
dimension grows. The efficiency gain at $d=50$ becomes very larger, while  LSM requires
$108.77$ seconds, RLSM only takes $1.93$ seconds. So here we start to
see the main strength of the method as RLSM keeps a level of price
and EE accuracy, while the computational cost grows
much less than for the benchmark. This is also fully consistent with the behaviour reported in
\citep{HerreraKrachRuyssenTeichmann2023}, even though the implementation used in
this article required some modifications to adapt it to exposure estimation from the configuration proposed in
that paper.

Another point is worth mentioning is that in the $d=50$ Black-Scholes case, LSM
already begins to show convergence problems, as the time-zero price is lower than the
European benchmark by a wider margin, while the initial exposure is
higher. This results suggest that the polynomial regression in LSM is starting to lose
accuracy at this dimension for the chosen path number, and in the next section we will analyze this further as it becomes more noticable.

\subsubsection{Heston Results}

The Heston results are given in Table~\ref{tab:maxcall_heston_results}. For the lower and intermediate dimensions, the two price estimates
remain close, at $d=5$, LSM gives $8.1276$ and RLSM gives $7.9415$, at $d=10$,
$11.4805$ and $11.3345$, and at $d=25$, $15.8409$ and $15.8387$. These are all
reasonable approximations to the corresponding European benchmark, and the
standard errors of both methods are now also closer to each other than in
the Black-Scholes case.

\begin{table}[H]
\centering
\begin{tabular}{lccccc}
\toprule
Model & $d$ & Price$_0$ & Euro MC & EE$(0)$ & Time (s) \\
\midrule
LSM  & 5  & 8.1276 (0.0464) & 8.1310 & 8.1724 & 1.84 \\
RLSM & 5  & 7.9415 (0.0582) & 8.1310 & 7.9670 & 0.52 \\
LSM  & 10 & 11.4805 (0.0469) & 11.6326 & 11.5986 & 4.53 \\
RLSM & 10 & 11.3345 (0.0516) & 11.6326 & 11.4202 & 0.89 \\
LSM  & 25 & 15.8409 (0.0427) & 16.3681 & 16.3819 & 42.97 \\
RLSM & 25 & 15.8387 (0.0682) & 16.3681 & 15.8924 & 1.82 \\
LSM  & 50 & 18.3571 (0.0777) & 19.8909 & 20.2777 & 392.62 \\
RLSM & 50 & 19.2335 (0.0608) & 19.8909 & 19.2948 & 2.91 \\
\bottomrule
\end{tabular}
\caption[American max-call results under Heston dynamics.]{American max-call results under Heston dynamics. Standard errors are reported in parentheses. Source: Own elaboration.}
\label{tab:maxcall_heston_results}
\end{table}

\begin{figure}[H]
\centering

\begin{subfigure}{0.45\textwidth}
    \centering
    \includegraphics[width=\linewidth]{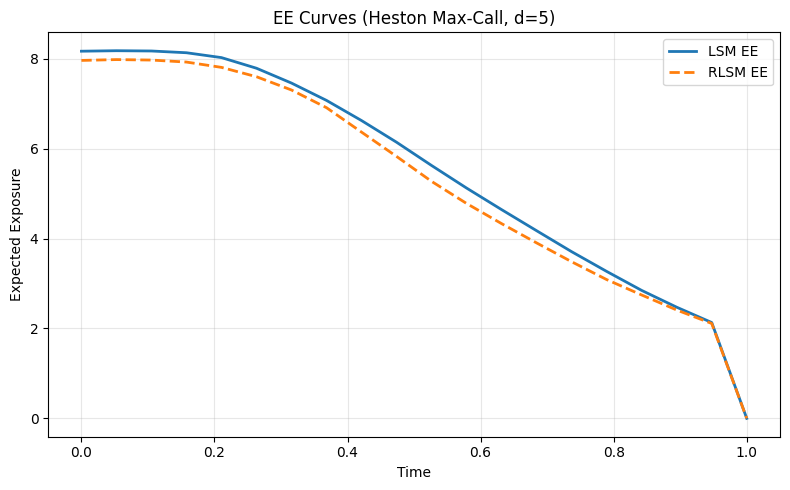}
    \caption{d = 5}
\end{subfigure}
\hfill
\begin{subfigure}{0.45\textwidth}
    \centering
    \includegraphics[width=\linewidth]{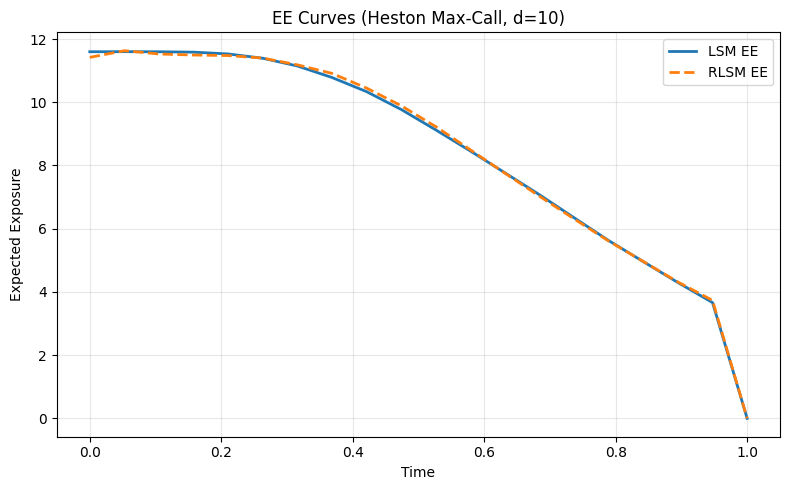}
    \caption{d = 10}
\end{subfigure}

\vspace{0.3cm}

\begin{subfigure}{0.45\textwidth}
    \centering
    \includegraphics[width=\linewidth]{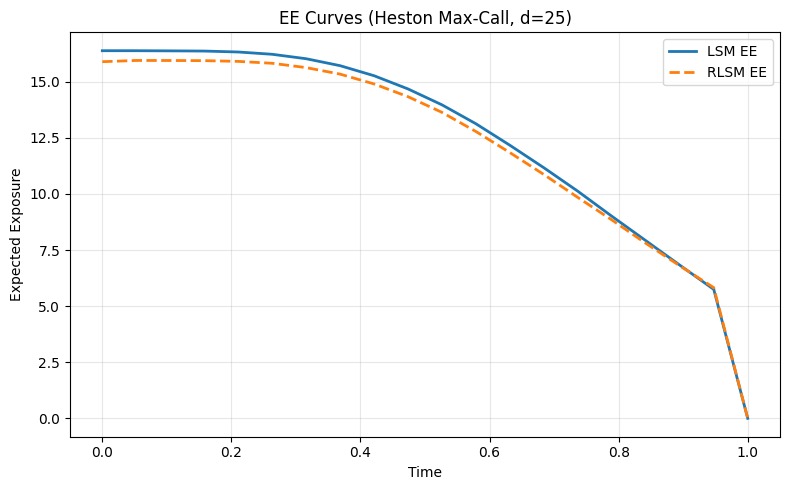}
    \caption{d = 25}
\end{subfigure}
\hfill
\begin{subfigure}{0.45\textwidth}
    \centering
    \includegraphics[width=\linewidth]{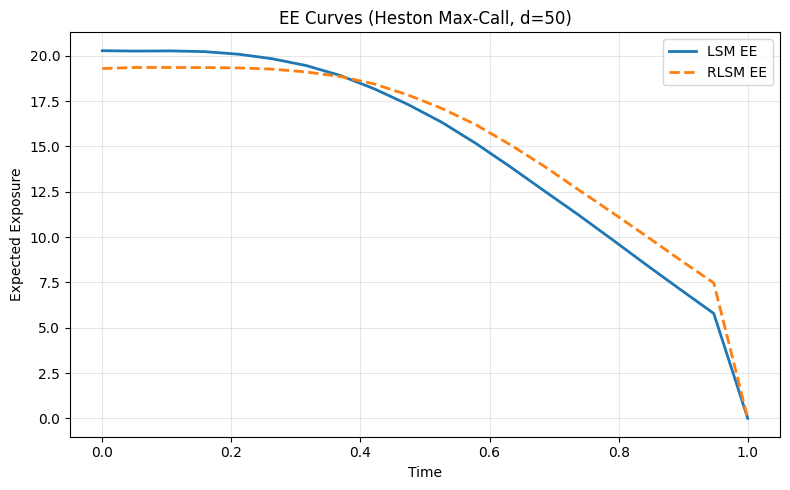}
    \caption{d = 50}
\end{subfigure}

\caption[EE curves for American max-call options under Heston dynamics across dimensions.]{EE curves for American max-call options under Heston dynamics across dimensions. Source: Own elaboration.}
\label{fig:maxcall_heston_ee}

\end{figure}

The EE curves in Figure~\ref{fig:maxcall_heston_ee} show that the convergence of
RLSM to LSM is again very good across the whole time horizon. In practice, the method remains very stable, while the computational
advantage over LSM becomes even larger.

That efficiency gain is once again even higher for high dimensionality than for Black-Scholes, the LSM runtime rises from $1.84$ seconds
at $d=5$ to $392.62$ seconds at $d=50$, while the corresponding RLSM runtimes
move from $0.52$ to $2.91$ seconds. Therefore, the same conclusion as in
Black-Scholes, but even more strong, once both the number of
underlyings and the number of risk factors increase, the classical regression in
LSM benchmark becomes increasingly expensive, while RLSM scales much better.

The $d=50$ Heston result shows better what we already discussed during Black-Scholes, as the LSM price
falls below the European benchmark, while the initial exposure is
substantially too high and the exposure curve differs more from RLSM that remains much closer to the reference
level in both price and initial EE. This discrepancy between the two methods at 
$d=50$ should not really be interpreted as RLSM not converging, because is the LSM benchmark the one that doesn't converges at the 10.000 path simulation to an accurate estimation.
This is exactly the sort of situation in which the scalability argument behind RLSM becomes practically relevant.

\begin{figure}[H]
\centering
\includegraphics[width=0.75\textwidth]{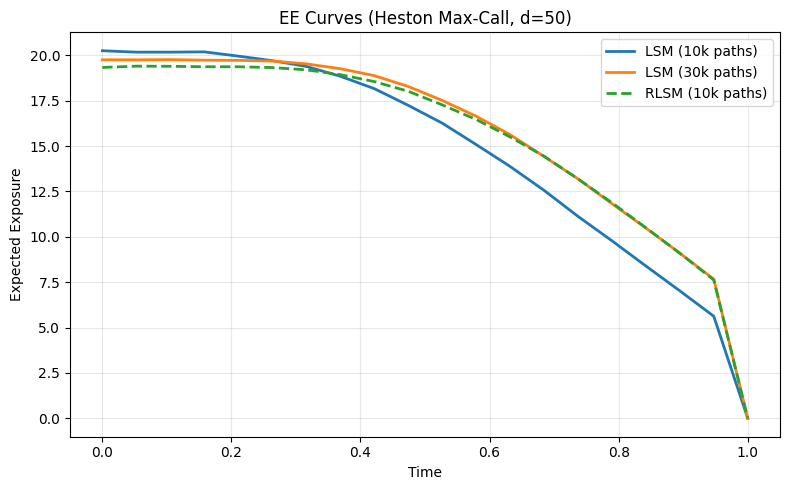}
\caption[EE comparison for the 50-dimensional Heston max-call.]{EE comparison for the 50-dimensional Heston max-call. LSM with 10.000 paths, LSM with 30.000 paths, and RLSM with 10.000 paths. Source: Own elaboration.}
\label{fig:maxcall_heston_d50_paths}
\end{figure}

To illustrate these convergence problems of LSM for high dimensions, an additional $d=50$ Heston comparison it's
shown in Figure~\ref{fig:maxcall_heston_d50_paths}. When LSM is ran with the
standard 10.000 paths it's EE profile still differ, but once the paths
are increased to 30.000, the LSM curve moves much closer to the one of RLSM, showing that the issue comes from regression benchmark rather than from the randomized
method. However for that number of paths the cost becomes even larger, the LSM runtime rises from about
$362.68$ seconds to $1070.88$ seconds.

This another reason of the advantage of RLSM, as the dimensions grows, the
regression problem becomes harder and LSM can still be improved by using more
paths, but doing so very quickly becomes expensive. In practice that means the
method loses efficiency precisely in the regime where one would most like to
preserve it. The analysis in Section \ref{path conv} returns to this issue more directly by
looking at path convergence itself.

\subsubsection{PFE Results}

Finally, we also consider PFE for the Black-Scholes max-call case, but since PFE is
not used later in the CVA application, the discussion here will remain brief. The
main here conclusion is the same as for EE, RLSM still converges well while remaining
much more efficient than LSM. At the same time, as for the vanilla option, because PFE is a tail quantity,
the LSM convergence issues become more visible there than in the mean exposure profile.
For that reason, the LSM benchmark in the PFE plot is run with 30.000 paths, as for only 10.000 paths, the gap would be larger, not because the RLSM tail
behaviour, but because LSM has still not fully converged in the
higher-dimensional case.

\begin{figure}[H]
\centering

\begin{subfigure}{0.45\textwidth}
    \centering
    \includegraphics[width=\linewidth]{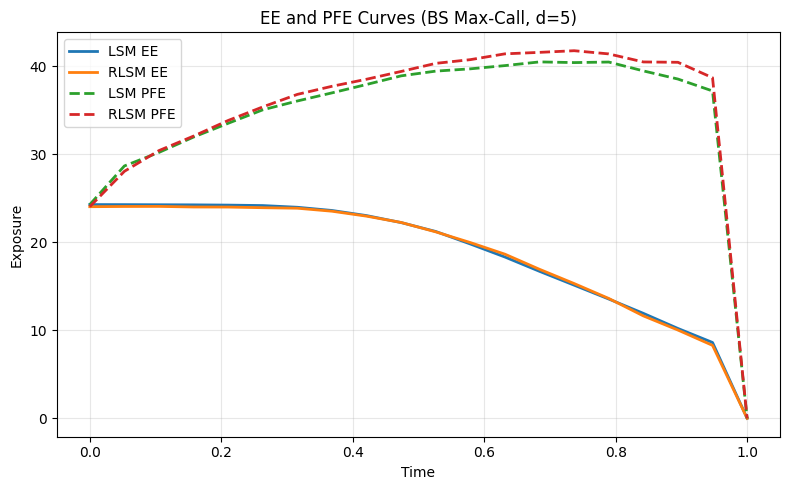}
    \caption{d = 5}
\end{subfigure}
\hfill
\begin{subfigure}{0.45\textwidth}
    \centering
    \includegraphics[width=\linewidth]{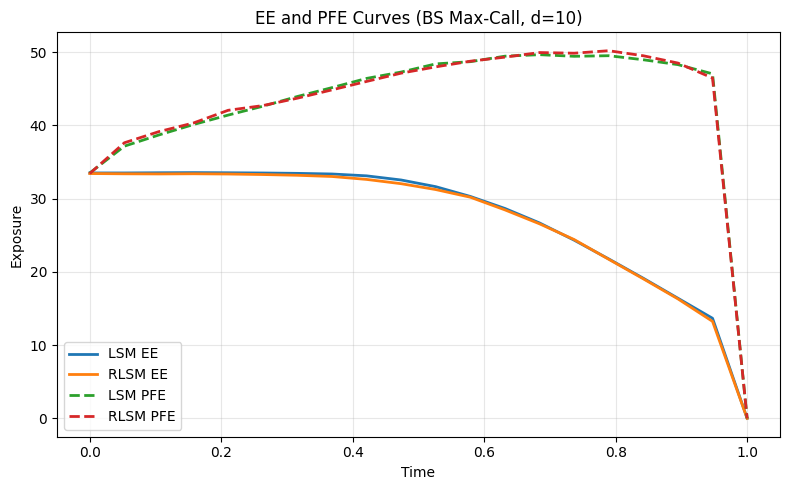}
    \caption{d = 10}
\end{subfigure}

\vspace{0.3cm}

\begin{subfigure}{0.45\textwidth}
    \centering
    \includegraphics[width=\linewidth]{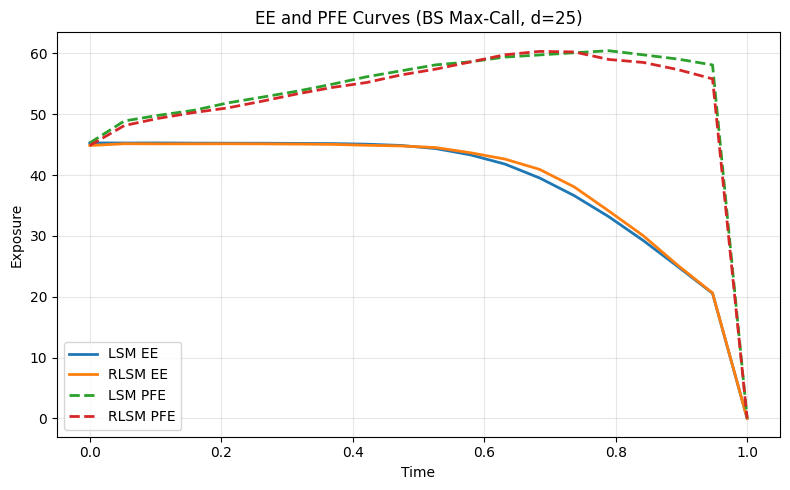}
    \caption{d = 25}
\end{subfigure}
\hfill
\begin{subfigure}{0.45\textwidth}
    \centering
    \includegraphics[width=\linewidth]{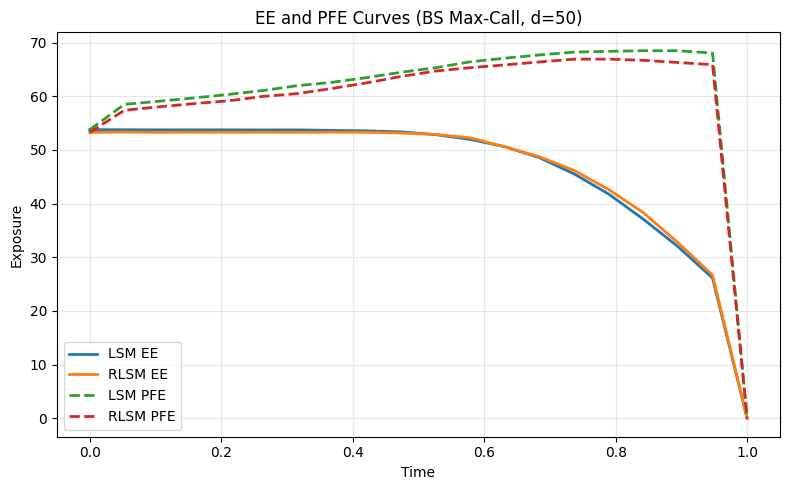}
    \caption{d = 50}
\end{subfigure}
\caption[PFE curve for American max-call options under Black-Scholes dynamics across dimensions.]{PFE curve for American max-call options under Black-Scholes dynamics across dimensions. Source: Own elaboration.}
\label{fig:maxcall_bs_pfe}
\end{figure}

In summary, the high-dimensional experiments show the different results than the vanilla case, here RLSM is the method that better preserves accuracy once the stopping problem becomes genuinely difficult, while having a lower computational cost. This is the main empirical result of this chapter, for low-dimensions the LSM benchmark is very efficient and accurate, while RLSM introduce extra noise and complexity, however for high dimensions this complexity becomes a feature that makes the model more robust and efficient than benchmark.

\subsection{Path Convergence} \label{path conv}

During this section we extend the previous comparison by looking directly at how both methods behave as the number of simulated paths change. The analysis is carried out on American max-call options under Heston dynamics, since this is the more realistic model and the one in which convergence problems become
more visible. The convergence is measured through the option price, using the
corresponding European price as a reference, because a very large LSM simulation as the benchmark is not feasible for the larger dimensions due to RAM limitations, so the European reference
is the more convenient choice here as in Section \ref{high dimension}.

\begin{figure}[H]
\centering
\includegraphics[width=0.60\textwidth]{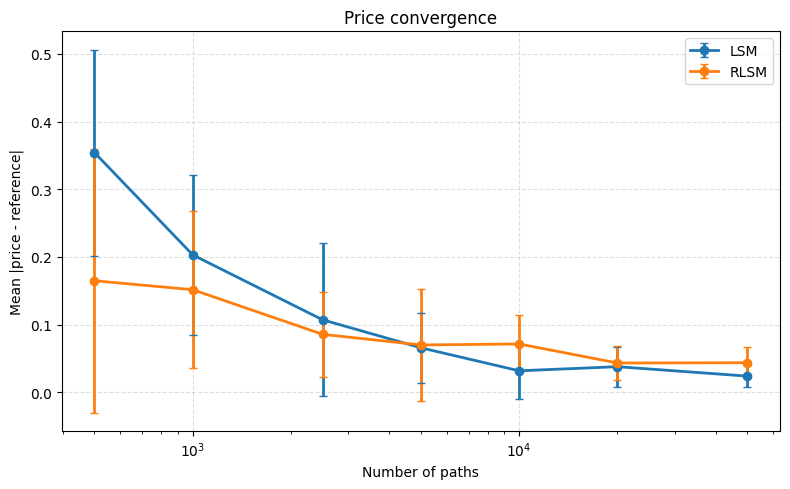}
\caption[Price convergence with respect to the number of paths for the 5-dimensional Heston max-call option.]{Price convergence with respect to the number of paths for the 5-dimensional Heston max-call option. Source: Own elaboration.}
\label{fig:path_conv_d5}
\end{figure}

The first plot, shown in Figure~\ref{fig:path_conv_d5}, corresponds to the 5-dimensional case. Here both methods behave in a similar way and even though RLSM appears to stabilize a little earlier when the number of paths are very small, the overall difference is not especially large. This is consistent with the discussion in the previous subsection, since five dimensions are still not enough for the scalability advantage of RLSM to become truly decisive.

\begin{figure}[H]
\centering
\includegraphics[width=0.60\textwidth]{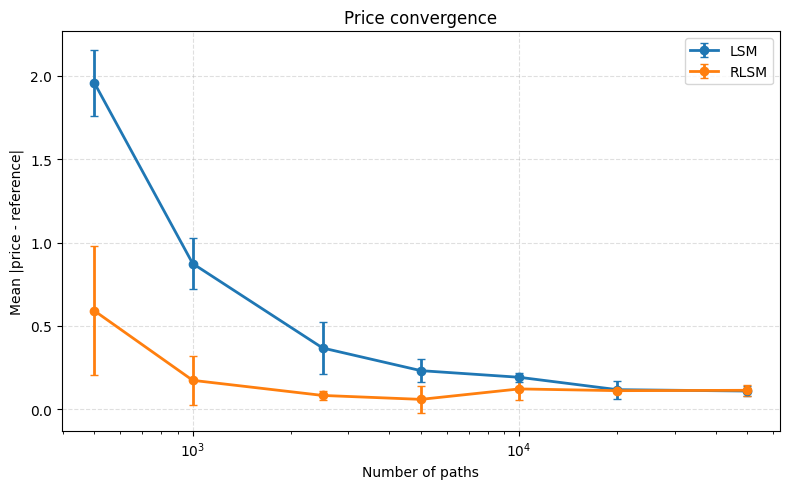}
\caption[Price convergence with respect to the number of paths for the 10-dimensional Heston max-call option.]{Price convergence with respect to the number of paths for the 10-dimensional Heston max-call option. Source: Own elaboration.}
\label{fig:path_conv_d10}
\end{figure}

The picture changes once we move to ten dimensions, as shown in Figure~\ref{fig:path_conv_d10}. At that point the difference is already much clearer, because RLSM reaches values close to its large-path estimate much earlier, while the LSM error becomes much larger when the path budget is small and only decreases once many more paths are used. So even before entering the really high-dimensional regime, one can already see that the classical regression becomes much more sensitive to the size of the Monte Carlo sample.

\begin{figure}[H]
\centering
\includegraphics[width=0.60\textwidth]{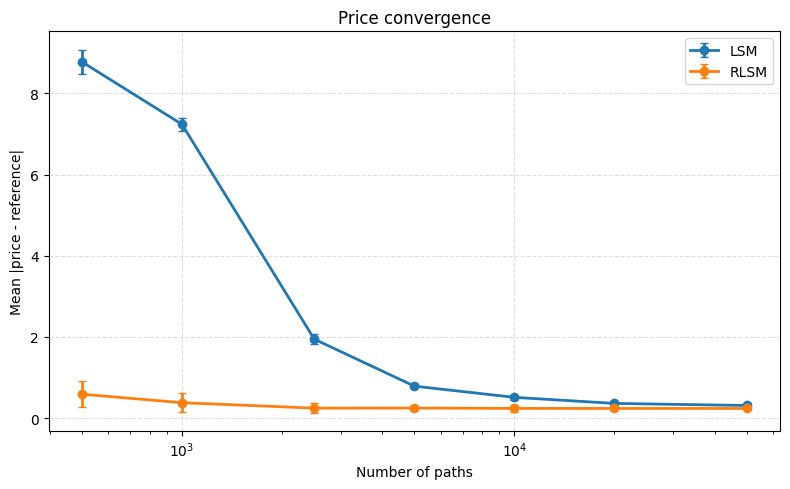}
\caption[Price convergence with respect to the number of paths for the 25-dimensional Heston max-call option.]{Price convergence with respect to the number of paths for the 25-dimensional Heston max-call option. Source: Own elaboration.}
\label{fig:path_conv_d25}
\end{figure}

This effect becomes stronger again in the 25-dimensional case, reported in Figure~\ref{fig:path_conv_d25}. Here the LSM error for low path numbers is so large that the RLSM convergence almost looks flat by comparison, simply because the scale of the plot is dominated by the LSM deviations. In practice, what this means is that the gap between both methods is no longer marginal, as the number of dimensions grows, LSM needs a larger number of paths before it starts to produce accurate estimates, while RLSM remains comparatively stable throughout the number of paths. This results are in line with what was already suggested in Section \ref{high dimension}, where the 50-dimensional results showed that LSM could still be brought closer to convergence, but only at a very high computational cost.

Overall, these plots confirm the same idea, when the problem remains relatively moderate, both methods can converge with a similar number of paths, even if RLSM still tends to stabilize a bit earlier. But as dimensionality increases, the difference becomes much larger, and the classical LSM regression becomes increasingly inaccurate, while RLSM reaches stable values with fewer simulations. At the same time, as the number of dimensions increases, RLSM is also fed with a richer set of features, which in general tends to be beneficial for neural-network-based methods, since they usually perform better when more relevant information is available. This is exactly the reason why the randomized approach becomes attractive in the high-dimensional case, not only because it dominates LSM in terms of costs, but also because its convergence remains more manageable once the regression problem becomes genuinely difficult.

\subsection{Sensitivity Analysis}

In this section we analyze the sensitivity of the main RLSM hyperparameters, including the number of neurons in the hidden layer, the ridge regularization parameter $\lambda$, and the input scaling factor. The goal here is to understand how the performance of the model changes when its key parameters are changed, and from there extract a connclusion about how to set the model parameters depending on the configuration of the problem.

\subsubsection{Hidden Size}

\begin{figure}[H]
\centering
\includegraphics[width=1.0\textwidth]{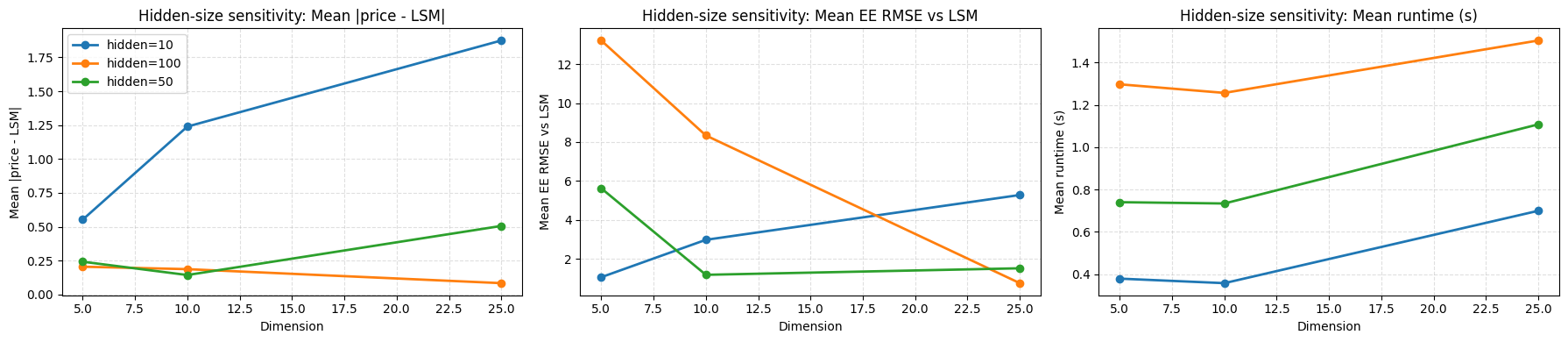}
\caption[Hidden-size sensitivity for the Black-Scholes model.]{Hidden-size sensitivity for the Black-Scholes model. Source: Own elaboration.}
\label{fig:bs_hidden_size}
\end{figure}

\begin{figure}[H]
\centering
\includegraphics[width=1.0\textwidth]{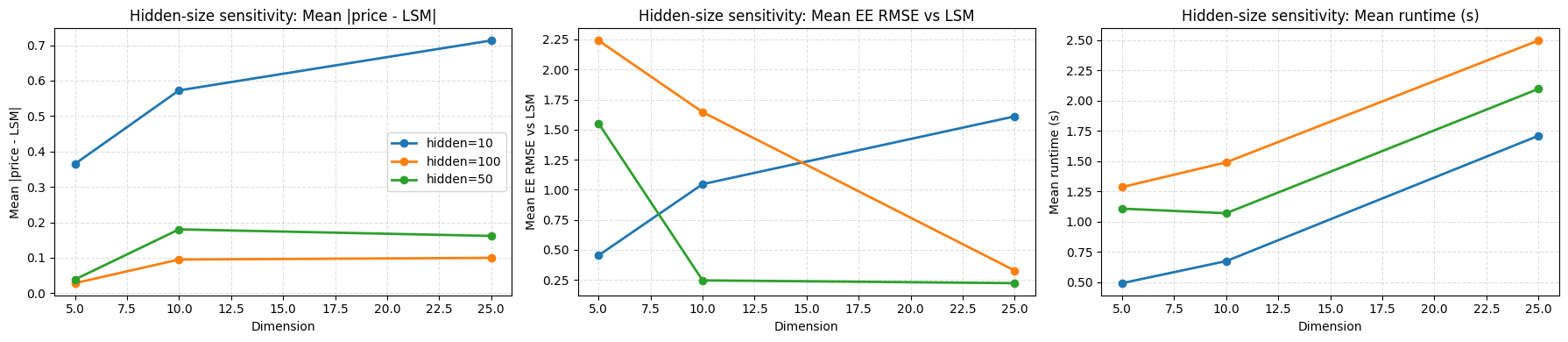}
\caption[Hidden-size sensitivity for the Heston model.]{Hidden-size sensitivity for the Heston model. Source: Own elaboration.}
\label{fig:heston_hidden_size}
\end{figure}

The analysis begins with the hidden size, that is the number of neurons in the randomized
hidden layer. Figures~\ref{fig:bs_hidden_size} and
\ref{fig:heston_hidden_size} report the corresponding sensitivity results under
Black-Scholes and Heston dynamics.

For the Black-Scholes model, the effect of the number of neurons is different
depending on whether one looks at pricing or at EE. In pricing, a larger hidden
layer generally tends to reduce the error, which is consistent with the fact
that the model often has a slightly downward bias with respect to LSM \citep{HerreraKrachRuyssenTeichmann2023}, so adding
more neurons helps raise the estimated price and reduces that gap. However for EE the results are 
different. For the lower dimensions, fewer neurons often
work better, while in the higher dimensions a larger hidden layer becomes give better results, 
as the number of inputs also increases and the
network needs enough capacity to process that additional information.

This difference between price and EE is also quite intuitive. Price is a single
observation at time zero, so moderate overfitting doesnt have big effects,
whereas EE depends on the whole exposure curve and is therefore more sensitive
to irregularities, peaks or lack of smoothness. In that sense, using fewer
neurons may act as a form of implicit regularization, this is something already mentioned in \citep{HerreraKrachRuyssenTeichmann2023}. 
Additionally an obvious computational trade-off, since hidden size is the
one parameter here that directly changes the size of the model and therefore has
a visible effect on runtime.

For Heston the conclusions are very similar. Pricing generally benefits from a
larger hidden layer, while for EE the preferred hidden size still depends on the
dimension, smaller values are enough in the lower-dimensional cases, whereas
higher dimensions usually need more neurons.

\subsubsection{Regularization Parameter}

\begin{figure}[H]
\centering
\includegraphics[width=1.0\textwidth]{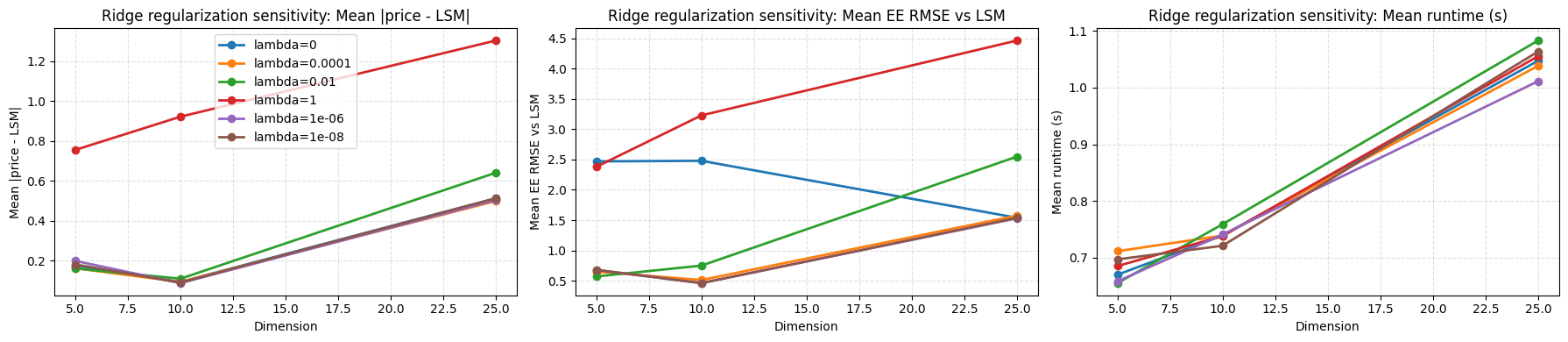}
\caption[Regularization sensitivity for the Black-Scholes model.]{Regularization sensitivity for the Black-Scholes model. Source: Own elaboration.}
\label{fig:bs_lambda}
\end{figure}

\begin{figure}[H]
\centering
\includegraphics[width=1.0\textwidth]{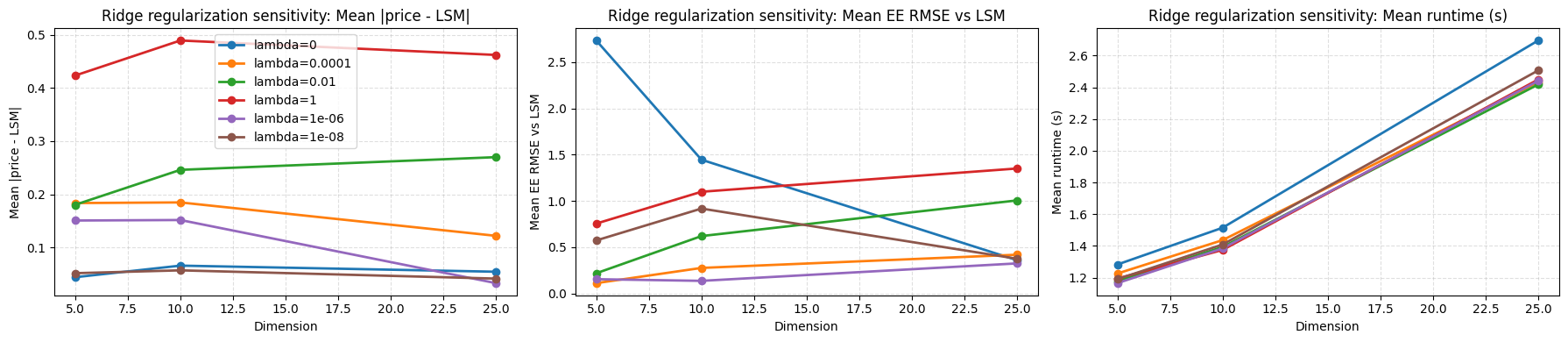}
\caption[Regularization sensitivity for the Heston model.]{Regularization sensitivity for the Heston model. Source: Own elaboration.}
\label{fig:heston_lambda}
\end{figure}

The ridge regularization parameter is considered next. Since the results
in Figures ~\ref{fig:bs_lambda} and ~\ref{fig:heston_lambda} are very similar for Black-Scholes and Heston we will
analyse them together.

For pricing, RLSM generally works best with low regularization values or even
with no regularization at all, while for EE the situation is different, since
both too much regularization and no regularization tend to give worse results,
whereas intermediate values, between $10^{-4}$ and $10^{-8}$, usually
produce the best behaviour. As in the previous subsection, the intuition is that
price is only a single observation, so the noise coming from the random
features or from a slight overfitting, does not affect it so much,
whereas for EE those same effects can generate peaks or curves that are less
smooth and therefore less accurate. Another important point is that the
regularization changes introduced in this article make even very small values of
$\lambda$ already meaningful, so the useful range appears earlier than one might
expect.

At the same time, regularization has no appreciable effect on runtime, since it
only changes the conditioning of the readout problem and not its overall size.
Still, as advised in the original paper \citep{HerreraKrachRuyssenTeichmann2023},
it is often preferable to use a reduced number of neurons as a form of
regularization, instead of taking a larger hidden layer that later requires more
ridge penalization, since reducing the number of neurons also reduces the
computational time and in many cases leads to very similar results.

\subsubsection{Input Scaling Factor}

\begin{figure}[H]
\centering
\includegraphics[width=1.0\textwidth]{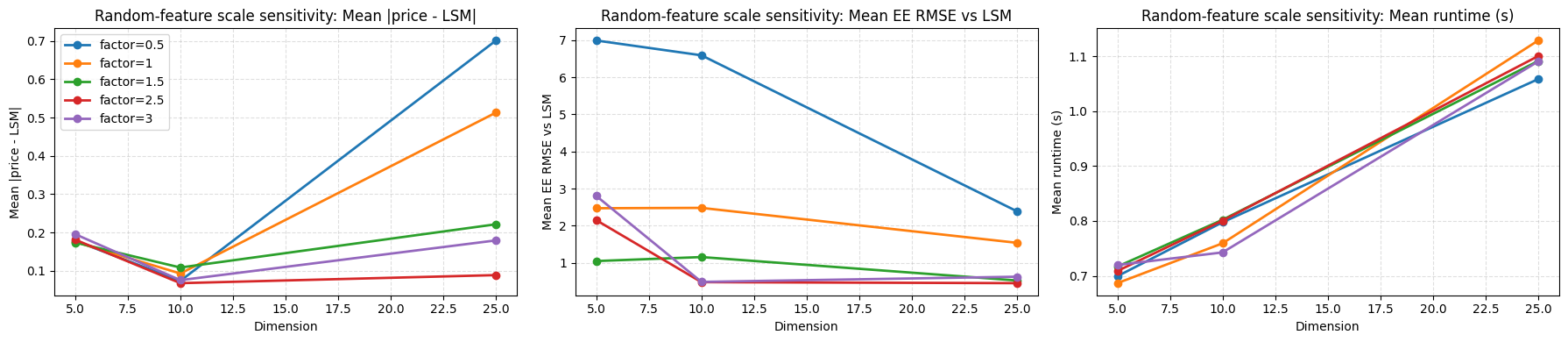}
\caption[Scaling factor sensitivity for the Black-Scholes model.]{Scaling factor sensitivity for the Black-Scholes model. Source: Own elaboration.}
\label{fig:bs_scaling}
\end{figure}

\begin{figure}[H]
\centering
\includegraphics[width=1.0\textwidth]{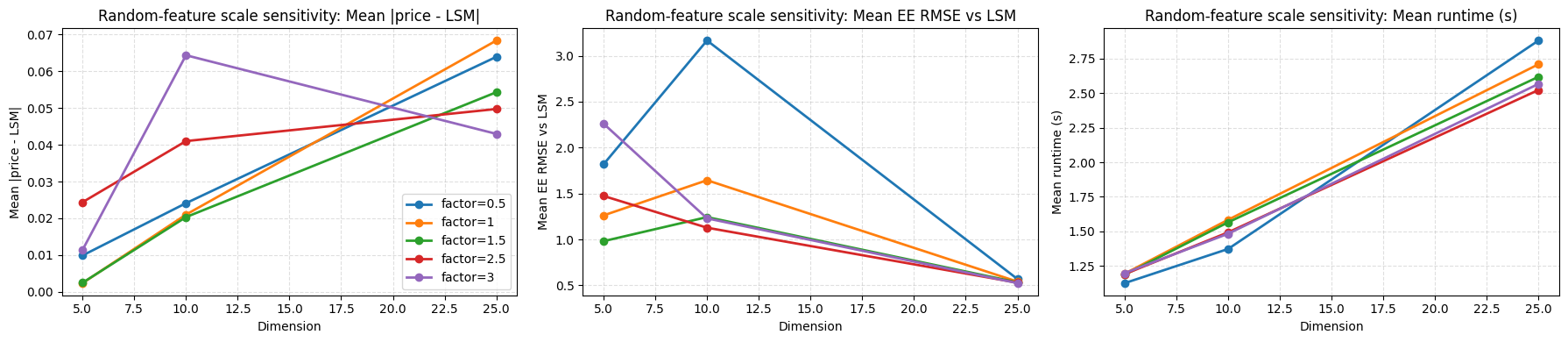}
\caption[Scaling factor sensitivity for the Heston model.]{Scaling factor sensitivity for the Heston model. Source: Own elaboration.}
\label{fig:heston_scaling}
\end{figure}

Finally, Figures ~\ref{fig:bs_scaling} and ~\ref{fig:heston_scaling} report the sensitivity of the
model with respect to the input scaling factor, again discusing Black-Scholes and
Heston together since the results are very similar in both cases.

The effect of this factor is harder to judge, mainly because it depends
quite strongly on the rest of the model configuration. Even so, a clear
pattern still appears in both market models, medium and high values tend to
work better in general, for both pricing and EE, while low values such as $0.5$
usually perform worse. A possible explanation is that a higher factor
amplifies the features, and therefore the changes in those features have a
larger influence on the randomized hidden layer.

\subsubsection{Conclusion}

To conclude this sensitivity analysis, the main point is that RLSM, like most machine
learning models, is sensitive to its hyperparameters, and therefore obtaining
the best performance requires either taking into account the empirical patterns
shown here or carrying out some parameter tuning for the specific problem under
consideration. At the same time, the analysis is useful because it also shows
that the model behaves in an interpretable way, hidden size have direct effect in all categories, regularization is especially relevant for exposure quality,
and the input factor matters, although in a more configuration-dependent manner.
It is also positive that the results for Black-Scholes and Heston are generally similar, since that points to a good degree of consistency in the model.

\clearpage
\section{CVA Case Study} \label{cva case study}

After conducting the empirical experiments comparing both models, the results suggest that RLSM is a good alternative to LSM for calculating exposure profiles, and therefore suitable for CVA estimation, because it preserves the convergence to the benchmark while offering a more efficient and scalable approximation in high-dimensional settings.
Additionally, another goal is to show how the individual exposure previously simulated behave in an aggregated
portfolio considering netting effects, and under different hazard rate configurations.

This portfolio consist in five American max-call options, including different levels of dimensionality and moneyness 
configurations (ITM, ATM and OTM), under a realistic shared 50 dimensional Heston paths simulation for both RLSM and LSM, so
that the case remains high-dimensional, as this is the right application for RLSM, while still reflects an actual 
counterparty credit risk problem.

\subsection{Case Study Design}

As previously mentioned, the netting set is composed of five american max-call options,
including three long ATM max-call options with
$10$, $25$ and $50$ underlying assets, one long OTM max-call with strike
$110$, and one short ITM max-call with strike $90$ and a notional
$0.75$, all of them with a intial spot of $100$ (Table~\ref{tab:cva_trade_spec}). The portfolio is design 
to include products previously tested, but including new configurations (ITM and OTM)
to confirm that the convergences of RLSM is not affected. Additionally, the role of the 
short options allows for the implementation of the netting effects, reducing the final
exposure of the portfolio. All the options are valued on the same shared 50 dimensional
Heston simulation, what allows us to aggregate the estimated values path by
path before taking the positive part, as is required for a netting set CVA calculation.

\begin{table}[H]
\centering
\begin{tabular}{lccccccc}
\toprule
Trade & Side & $d$ & Moneyness & Strike & Notional & Assets \\
\midrule
ATM-10D & Long  & 10 & ATM & 100 & 1.00 & 1-10 \\
ATM-25D & Long  & 25 & ATM & 100 & 1.00 & 1-25 \\
ATM-50D & Long  & 50 & ATM & 100 & 1.00 & 1-50 \\
OTM-10D & Long  & 10 & OTM & 110 & 1.00 & 11-20 \\
ITM-10D-Short & Short & 10 & ITM & 90 & 0.75 & 21-30 \\
\bottomrule
\end{tabular}
\caption[Trade specification of the five-trade American max-call netting set.]{Trade specification of the five-trade American max-call netting set. Source: Own elaboration.}
\label{tab:cva_trade_spec}
\end{table}

For the credit calculations, three deterministic hazard-rate scenarios are used,
\emph{Low}, \emph{Base} and \emph{Stress} (Table~\ref{tab:cva_credit_scenarios}). The recovery rate is assumed to be $R=40\%$ 
for all of them, to show how the portfolio CVA reacts to changes in credit quality
without turning the case study into a separate calibration exercise.

\begin{table}[H]
\centering
\captionsetup{justification=centering}
\begin{tabular}{lcc}
\toprule
Scenario & Hazard Rates & Recovery \\
\midrule
Low    & [0.006, 0.007, 0.008, 0.009] & 0.40 \\
Base   & [0.015, 0.017, 0.019, 0.021] & 0.40 \\
Stress & [0.035, 0.040, 0.045, 0.050] & 0.40 \\
\bottomrule
\end{tabular}
\caption[Credit scenarios used in the CVA case study.]{Credit scenarios used in the CVA case study. 
The hazard rates correspond to the intervals $[0,0.25]$, $[0.25,0.50]$, $[0.50,0.75]$, and $[0.75,1.00]$ years. Source: Own elaboration.}

\label{tab:cva_credit_scenarios}
\end{table}

Regarding the number of paths, LSM is run with 30,000 paths, whereas RLSM uses 10,000 paths. 
As already shown in Section \ref{high dimension}, the 50D Heston LSM
benchmark was not yet properly converged at the specified number of paths, so using the same
number of paths for both methods would have meant computing the LSM CVA from an
under-converged exposure profile. Since this is a case study the objective of
LSM is to provide a real benchmark, and that requires giving enough paths to ensure that the 
results of this benchmark are accurate. 

\subsection{Trade-Level Results}

\begin{figure}[H]
\centering
\includegraphics[width=0.95\textwidth]{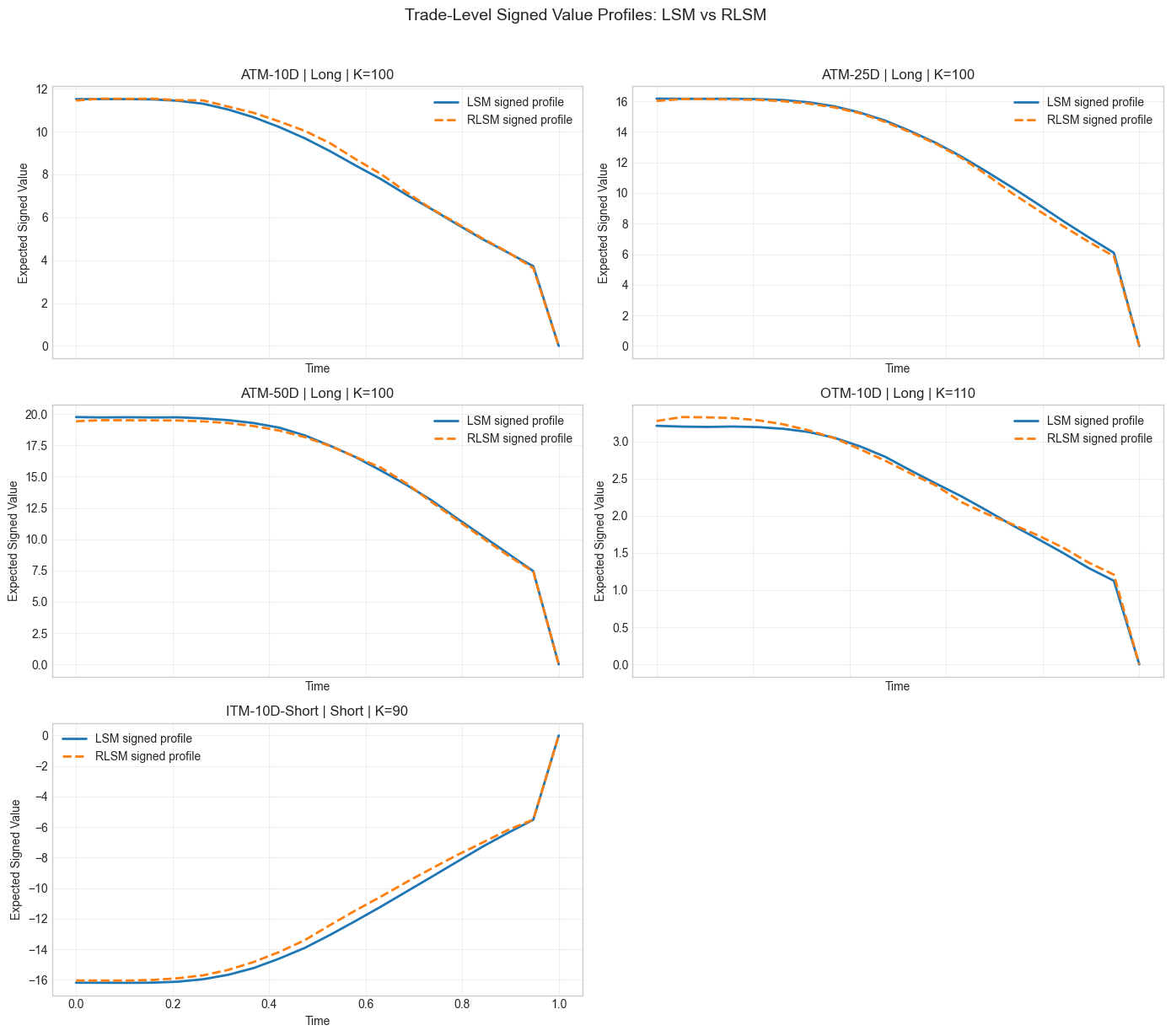}
\caption[Trade-level option values profiles for the five-trade Heston netting set.]{Trade-level option values profiles for the five-trade Heston netting set. Source: Own elaboration.}
\label{fig:cva_trade_signed_profiles}
\end{figure}

Before moving to the real case study where we analyze the portfolio CVA, it is useful to check whether both methods
produce similar trade-level profiles, since these are the values that are
later aggregated into the netting set. In Figure~\ref{fig:cva_trade_signed_profiles}
this profiles are observed for all the products within the portfolio, and a very similar results are observed
for LSM and RLSM.

Additionally, the the ITM short position has a negative profile, meaning the its exposure
for the individual unilateral CVA is 0. However, later on for the portfolio case, once it is
netted in the portfolio, it will offset the positive value of the 
long options, and therefore reduce the final exposure of the overall portfolio. Thats the reason
why the sum of individual trade unilateral CVA differs from the portfolio unilateral CVA.

In Table~\ref{tab:cva_trade_level} trade-level CVA results are shown. The largest
standalone CVA contribution comes from the 50D ATM option,
followed by the 25D ATM option, while the OTM position remains much smaller and
the short ITM trade contributes zero unilateral CVA, as previously mentioned. 
Already at trade level, the case study shows that the exposure profiles
estimated by RLSM lead to practically the same unilateral CVA results as
those produced by the LSM benchmark.

\begin{table}[H]
\centering
\begin{tabular}{lcccccc}
\toprule
Trade & Scenario & LSM CVA & LSM Error & RLSM CVA & RLSM Error \\
\midrule
ATM-10D & Low    & 0.037966 & 0.000088 & 0.038484 & 0.000147 \\
ATM-10D & Base   & 0.091387 & 0.000206 & 0.092622 & 0.000346 \\
ATM-10D & Stress & 0.213175 & 0.000480 & 0.216059 & 0.000807 \\
ATM-25D & Low    & 0.057131 & 0.000103 & 0.056432 & 0.000165 \\
ATM-25D & Base   & 0.137306 & 0.000241 & 0.135665 & 0.000388 \\
ATM-25D & Stress & 0.320295 & 0.000563 & 0.316469 & 0.000904 \\
ATM-50D & Low    & 0.070707 & 0.000120 & 0.070131 & 0.000183 \\
ATM-50D & Base   & 0.169889 & 0.000280 & 0.168493 & 0.000428 \\
ATM-50D & Stress & 0.396306 & 0.000654 & 0.393050 & 0.001000 \\
OTM-10D & Low    & 0.010873 & 0.000040 & 0.011003 & 0.000061 \\
OTM-10D & Base   & 0.026155 & 0.000094 & 0.026472 & 0.000143 \\
OTM-10D & Stress & 0.061011 & 0.000219 & 0.061748 & 0.000334 \\
ITM-10D-Short & Low    & 0.000000 & 0.000000 & 0.000000 & 0.000000 \\
ITM-10D-Short & Base   & 0.000000 & 0.000000 & 0.000000 & 0.000000 \\
ITM-10D-Short & Stress & 0.000000 & 0.000000 & 0.000000 & 0.000000 \\
\bottomrule
\end{tabular}
\caption[Trade-level unilateral CVA results across credit scenarios.]{Trade-level unilateral CVA results across credit scenarios. Source: Own elaboration.}
\label{tab:cva_trade_level}
\end{table}

\subsection{Portfolio Exposure and CVA}

\begin{figure}[H]
\centering
\includegraphics[width=0.85\textwidth]{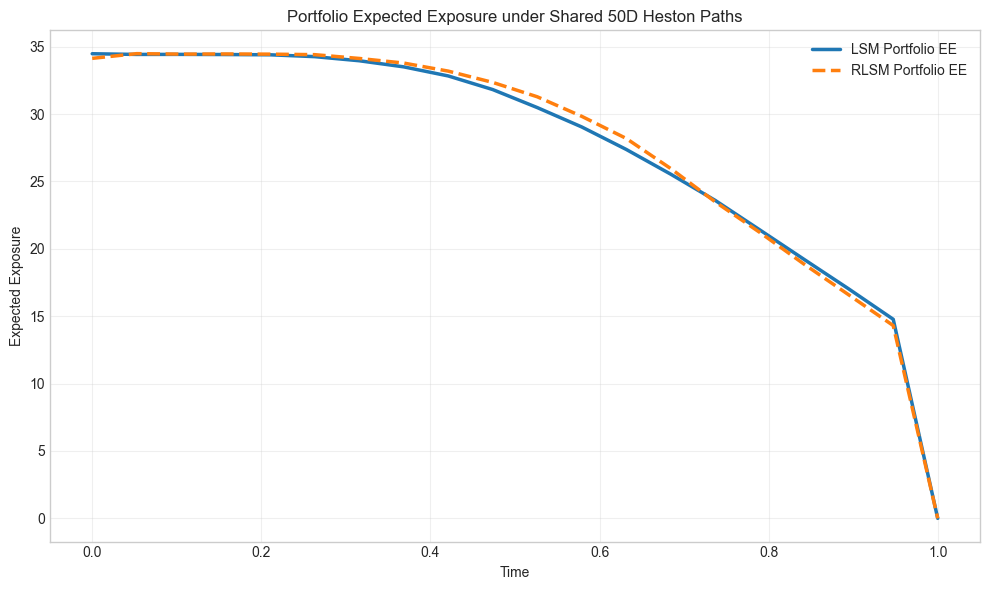}
\caption[Portfolio expected exposure under shared 50-dimensional Heston paths.]{Portfolio expected exposure under shared 50-dimensional Heston paths. Source: Own elaboration.}
\label{fig:cva_portfolio_ee}
\end{figure}

Moving into the portfolio CVA estimation, the first step is to aggregated all the paths
of the trades profiles to obtained the portolio EE curved already with the netting benefit included.
This Expected Exposure curve is shown in \ref{fig:cva_portfolio_ee}. Once again, exposure of both models
converges very close.  

This convergence is important because in the portfolio aggregation is where small 
differences in each trade could result in a larger difference in the portfolio level, 
especially considering the short position. However this is not observed, supporting the 
use of RLSM for portolio exposures under netting scenarios as the results converges to the
LSM benchmark.

The results of the portfolio CVA are summarized in
Table~\ref{tab:cva_portfolio_summary}. Along all three credit scenarios the results of LSM and
RLSM are extremely close. In the base scenario the portfolio CVA is
$0.300778$ under LSM and $0.302171$ under RLSM, while under stress the figures
rise to $0.701625$ and $0.704884$. The same happens with the standalone sums and
with the netting benefits, which remain very similar across both methods. For the base scenario, 
the netting effect is $0.123959$ for LSM and $0.121081$ for
RLSM, while under stress the benefit grows to $0.2892$ and $0.2824$,
respectively. So the two methods are not only close in the portfolio CVA itself,
but also the offsetting between trades are almost the same amount.

\begin{table}[H]
\centering
\small
\setlength{\tabcolsep}{4pt}
\begin{tabular}{lcccccccc}
\toprule
& \multicolumn{4}{c}{LSM} & \multicolumn{4}{c}{RLSM} \\
\cmidrule(lr){2-5} \cmidrule(lr){6-9}
Scenario & Standalone & Portfolio & Error & Netting & Standalone & Portfolio & Error & Netting \\
\midrule
Low    & 0.176676 & 0.125255 & 0.000232 & 0.051421 & 0.176050 & 0.125832 & 0.000366 & 0.050218 \\
Base   & 0.424737 & 0.300778 & 0.000544 & 0.123959 & 0.423252 & 0.302171 & 0.000859 & 0.121081 \\
Stress & 0.990786 & 0.701625 & 0.001270 & 0.289162 & 0.987326 & 0.704884 & 0.002004 & 0.282442 \\
\bottomrule
\end{tabular}
\caption[Portfolio CVA, error, and netting benefit across the three credit scenarios.]{Portfolio CVA, error, and netting benefit across the three credit scenarios. Source: Own elaboration.}
\label{tab:cva_portfolio_summary}
\end{table}

\begin{figure}[H]
\centering
\includegraphics[width=0.95\textwidth]{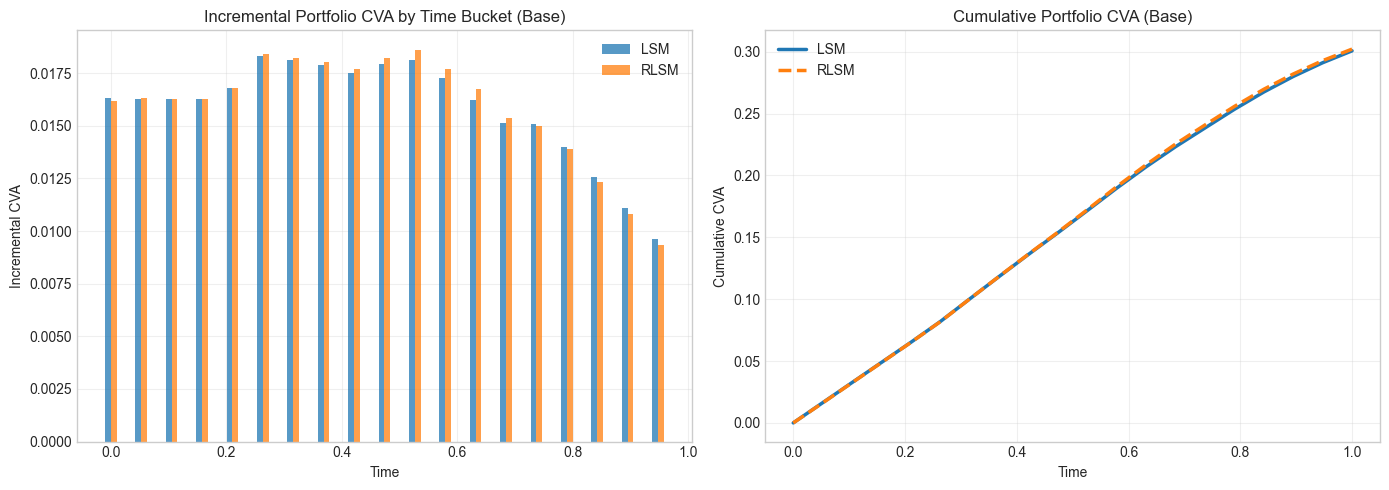}
\caption[Incremental and cumulative portfolio CVA in the base credit scenario.]{Incremental and cumulative portfolio CVA in the base credit scenario. Source: Own elaboration.}
\label{fig:cva_incremental_cumulative}
\end{figure}

The Figure~\ref{fig:cva_incremental_cumulative} gives a better view
of the base scenario along the time. Both the incremental CVA bars and the cumulative curves
are almost identical for both models over most of the horizon. This indicates that
the convergence between both CVA estimates is not just in the final aggregate number,
but it holds through the whole time decomposition of the exposure profile.

Finally regarding the computational time, the common 50D Heston path generation process, 
used for both models, took $12.31$ seconds. For the exposure estimation, 
while LSM is heavily affected by the high dimensionality and the need for more paths
to achieve accurate results, the efficiency observed during Chapter 6 held for RLSM, resulting
in runtimes of $822.37$ for LSM and $7.58$ seconds for RLSM.

\subsection{Case Study Conclusion}

The conclusion of this case study is that RLSM still produces robust results under 
a realistic unilateral CVA setting, including portfolio aggregation, netting effects
and shared market scenarios. RLSM was able to produce almost identical results of
portfolio EE, trade-level CVA, portfolio CVA and netting benefits compared with the
LSM benchmark. Additionally, the computational efficiency of RLSM remained as its 
main advantage, delivering comparable results while doing it more than 100 times 
faster than the LSM benchmark.

\clearpage
\section*{Conclusions} \label{conclusions}
\addcontentsline{toc}{section}{Conclusion}

Finally, we will conclude with the main findings obtained in this article and answering the research questions formulated in the introduction. The empirical evidence supports the idea that the randomized neural network approach can be extended from pricing to the estimation of exposure profiles and unilateral CVA while still converging to the LSM benchmark.

The results showed that for low dimensional products the LSM benchmark remains as a better choice because in this scenarios the LSM is still more efficient than RLSM while being more simple, easier to interpret, having fewer parameters, and producing accurate prices and stable exposure results. Meanwhile, for the high-dimensional setting, the randomized neural network approach shows a clearer advantage since the exposure profiles converge to the LSM benchmark, while the computational cost is lower and scales better, and this same convergence is preserved when those exposure profiles are aggregated in the portfolio used to calculate the unilateral CVA with netting effects considered in the case study. However, the higher complexity of the randomized neural networks also means that its performance depends on hyperparameter tuning, making the implementation more difficult and increasing the model risk, something especially relevant for highly regulated areas as Counterparty Credit Risk. Therefore, this model should be considered as an alternative for the scenarios where LSM becomes inefficient and unreliable.

Overall, the main conclusion of the article is that the randomized neural network approach proposed in \citep{HerreraKrachRuyssenTeichmann2023} can be extended from pricing to exposure and CVA estimation while keeping the convergence with respect to the LSM benchmark, and its main practical benefit appears in the high-dimensional cases, while LSM remains as the best model for the simpler low-dimensional ones.

Several limitations of this study should be acknowledged. The first one concerns the scope of the products and models considered, because the empirical analysis is restricted to American equity options, specifically vanilla products and max-call options, under Black-Scholes and Heston dynamics. Another limitation is that the methodological comparison is restricted to one randomized neural network specification, while \citep{HerreraKrachRuyssenTeichmann2023} several machine-learning models are compared. In this article only the specification that best fits the objective of extending the framework from pricing to exposure and CVA is implemented, making the comparison easier, but at the same time it means that the conclusions are tied to that specific randomized neural network design and not to the whole family of models considered by \citep{HerreraKrachRuyssenTeichmann2023}.

The experimental range is also limited by the available computational resources, specially by the RAM limitations for the simulations, becoming a direct constraint when trying to run experiments with larger dimensions or with much higher path numbers, since both the path generation and the storage of the simulated values become heavier as the size of the problem increases. Finally, the CVA case study is restricted to a unilateral framework with deterministic hazard-rate scenarios and portfolio netting, while other characteristics as collateral, wrong-way risk, DVA, funding effects and the broader xVA interactions are not included.

These limitations also suggest several directions for future research. A first extension could be to broaden the methodological comparison within the randomized framework proposed by \citep{HerreraKrachRuyssenTeichmann2023}. In particular, the Randomized Recurrent Least-Squares Monte Carlo (RRLSM) method could be implemented and compared with the RLSM specification used in this article. Other related approaches from the same family, such as Randomized Fitted Q-Iteration (RFQI), could also be considered in order to study whether the advantages observed here are specific to the least-squares formulation or remain present across other randomized neural-network methods.

Future work could also test the methodology under more demanding underlying dynamics. The empirical analysis in this article is restricted to Black-Scholes and Heston models, but further research could consider stochastic local volatility, stochastic volatility with jumps, or rough-volatility models. These settings may generate more complex continuation-value functions, making them an even better environment for randomized neural-network features, and offer a stronger advantage over the classical method.

The product scope could also be expanded to more complex early-exercise derivatives, such as Bermudan swaptions or callable exotic structures, where the exercise policy may be more non-linear than in the current products. Beyond prices and exposure profiles, future research could also study price and CVA sensitivities, in order to assess whether the same framework can be used not only for valuation and exposure estimation, but also for risk-management quantities. Finally, the CVA component could be extended toward a broader xVA setting, including collateralization, wrong-way risk, DVA, funding effects, or other valuation adjustments, in order to evaluate whether the computational advantages observed for RLSM remain useful in a more realistic counterparty-risk framework.

\bigskip

\noindent\textit{Source code}

\smallskip
\noindent\hspace*{1cm} The complete code can be found and downloaded from the GitHub repository:\\ 
\hspace*{1.2cm}\url{https://github.com/isidromoroso/thesis_final_code}

\clearpage
\addcontentsline{toc}{section}{References}
\bibliographystyle{apalike}
\bibliography{references}

\clearpage
\addcontentsline{toc}{section}{\listtablename}
\listoftables 

\addcontentsline{toc}{section}{\listfigurename} 
\listoffigures

\section*{List of Abbreviations}
\addcontentsline{toc}{section}{List of Abbreviations}

\begin{tabular}{ll}
\textbf{AMC} & American Monte Carlo \\
\textbf{CCR} & Counterparty Credit Risk \\
\textbf{ColVA} & Collateral Valuation Adjustment \\
\textbf{CVA} & Credit Valuation Adjustment \\
\textbf{DVA} & Debt Valuation Adjustment \\
\textbf{EE} & Expected Exposure \\
\textbf{ENE} & Expected Negative Exposure \\
\textbf{EPE} & Expected Positive Exposure \\
\textbf{FVA} & Funding Valuation Adjustment \\
\textbf{KVA} & Capital Valuation Adjustment \\
\textbf{LGD} & Loss Given Default \\
\textbf{LSM} & Least-Squares Monte Carlo \\
\textbf{MtM} & Mark-to-Market \\
\textbf{MVA} & Margin Valuation Adjustment \\
\textbf{OTC} & Over-the-Counter \\
\textbf{PD} & Probability of Default \\
\textbf{PFE} & Potential Future Exposure \\
\textbf{RLSM} & Randomized Least-Squares Monte Carlo \\
\textbf{xVA} & X-Valuation Adjustment \\
\end{tabular}
\newpage

\end{document}